\RequirePackage{ifpdf}
\ifpdf 
\documentclass[pdftex]{sigma}
\else
\documentclass{sigma}
\fi

\usepackage{mathrsfs}
\usepackage{eucal}
\usepackage{epsf}

\def\rank{\mathrm{rank\,}}
\def\tr{\mathrm{tr\,}}

\def\im{\mathrm{Im\,}}
\def\ad{\mathrm{ad\,}}

\def\openone{\leavevmode\hbox{\small1\kern-3.3pt\normalsize1}}

\def\a{{\boldsymbol a}}
\def\m{{\boldsymbol m}}
\def\b{{\boldsymbol b}}
\def\c{{\boldsymbol c}}
\def\d{{\boldsymbol d}}

\def\q{{\boldsymbol q}}

\def\h{{\boldsymbol h}}
\def\T{{\boldsymbol T}}
\def\S{{\boldsymbol S}}

\def\bPsi{{\boldsymbol \Psi}}
\def\bPhi{{\boldsymbol \Phi}}

\def\diag{\mbox{diag\,}}

\def\Res{\mathop{\mbox{Res}\,}\limits}

\def\newpic#1{%
   \def\emline##1##2##3##4##5##6{%
      \put(##1,##2){\special{em:point #1##3}}%
      \put(##4,##5){\special{em:point #1##6}}%
      \special{em:line #1##3,#1##6}}}

\newpic{}

\def\ad{\mbox{ad}\,}

\def\bbbe{\Bbb E}
\def\bbbr{{\Bbb R}}

\def\bbbc{{\Bbb C}}
\def\bbbe{{\Bbb E}}
\def\bbbz{{\Bbb Z}}

\def\bbbe{{\Bbb E}}
\def\bbbc{{\Bbb C}}
\def\bbbr{{\Bbb R}}

\def\bbbz{{\Bbb Z}}

\def\bbbz{{\Bbb Z}}
\newcommand{\rd}{\mathrm{d}} 
\newcommand{\re}{\mathrm{e}} 
\newcommand{\ri}{\mathrm{i}} 
\def\openone{\leavevmode\hbox{\small1\kern-3.3pt\normalsize1}}
\def\Res{\mathop{\mbox{Res}\,}\limits}

\def\ad{{\mbox{ad}}}

\allowdisplaybreaks
\def\tr{\mathrm{tr\,}}

\def\ad{\mathrm{ad\,}}

\def\tr{\mathrm{tr\,}}

\def\ad{\mathrm{ad\,}}

\def\openone{\leavevmode\hbox{\small1\kern-3.3pt\normalsize1}}

\def\a{{\boldsymbol a}}
\def\b{{\boldsymbol b}}
\def\c{{\boldsymbol c}}
\def\d{{\boldsymbol d}}

\def\T{{\boldsymbol T}}
\def\S{{\boldsymbol S}}

\def\0{{\boldsymbol 0}}
\def\bPsi{{\boldsymbol \Psi}}
\def\bPhi{{\boldsymbol \Phi}}

\def\ad{\mbox{ad\,}}

\def\tr{\mbox{tr\,}}
\def\diag{\mbox{diag\,}}

\def\bbbe{{\mathbb E}}
\def\bbbc{{\mathbb C}}
\def\bbbr{{\mathbb R}}

\begin{document}

\allowdisplaybreaks

\renewcommand{\thefootnote}{$\star$}

\renewcommand{\PaperNumber}{044}

\FirstPageHeading

\numberwithin{equation}{section}


\ShortArticleName{MNLS Models
on Symmetric Spaces: Spectral  and Representations Theory}

\ArticleName{Multi-Component NLS Models on Symmetric Spaces:\\ Spectral Properties versus Representations Theory\footnote{This
paper is a contribution to the Proceedings of the Eighth
International Conference ``Symmetry in Nonlinear Mathematical
Physics'' (June 21--27, 2009, Kyiv, Ukraine). The full collection
is available at
\href{http://www.emis.de/journals/SIGMA/symmetry2009.html}{http://www.emis.de/journals/SIGMA/symmetry2009.html}}}

\Author{Vladimir S. GERDJIKOV~$^\dag$ and Georgi G. GRAHOVSKI~$^{\dag\ddag}$}

\AuthorNameForHeading{V.S. Gerdjikov and G.G. Grahovski}

\Address{$^\dag$~Institute for Nuclear Research and Nuclear
Energy, Bulgarian Academy of Sciences,\\
$\phantom{^\dag}$~72 Tsarigradsko chaussee, 1784 Sofia, Bulgaria}

\EmailD{\href{mailto:gerjikov@inrne.bas.bg}{gerjikov@inrne.bas.bg}, \href{mailto:grah@inrne.bas.bg}{grah@inrne.bas.bg}}

\Address{$^\ddag$~School of Mathematical Sciences, Dublin Institute of Technology,\\
$\phantom{^\ddag}$~Kevin Street, Dublin 8, Ireland}
\EmailD{\href{mailto:georgi.grahovski@dit.ie}{georgi.grahovski@dit.ie}}

\ArticleDates{Received January 20, 2010, in f\/inal form May 24, 2010;  Published online June 02, 2010}

\Abstract{The algebraic structure and the spectral properties of a special class of multi-component NLS equations, related
to the  symmetric spaces of {\bf BD.I}-type  are analyzed. The focus of the study is on the spectral theory of the relevant
Lax operators for dif\/ferent fundamental representations of the underlying simple Lie algebra $ \mathfrak{g}$. Special attention is
paid to the structure of the dressing factors in spinor representation of the orthogonal simple Lie algebras of ${\bf B}_r\simeq so(2r+1, \bbbc)$ type.}


\Classification{37K20; 35Q51; 74J30; 78A60}

\section{Introduction}

The nonlinear Schr\"odinger  equation \cite{ZaSha1,APT}
\begin{gather*}
{\rm i}q_t + q_{xx} +2|q|^2q=0, \qquad q=q(x,t)
\end{gather*}
has natural multi-component generalizations. The f\/irst multi-component NLS type model with applications to
physics is the so-called vector NLS equation (Manakov model)~\cite{Ma1,APT}:
\begin{gather*}
i{\bf v}_t+{\bf v}_{xx}+2 ({\bf v}^\dag , {\bf v}) {\bf v}=0,\qquad {\bf v}=
\left(\begin{array}{c} v_1(x,t)\\ \vdots \\ v_n(x,t)\\ \end{array}\right).
\end{gather*}
Here ${\bf v}$ is an $n $-component complex-valued vector and $(\cdot ,\cdot) $ is the standard scalar product. All these models appeared to be integrable by the inverse scattering method \cite{AKNS,APT,ZMNP,FaTa,CaDeB,CaDe,GVY*08,KN79}.

Later on, the applications of the dif\/ferential geometric and Lie algebraic methods to soliton type equations  \cite{DrSok,Mikh,BeSat,vgn,vgrn,LoMi1,GGK05a,GGK05b,manev04,kagg07,gkv08,CaDe2,D1} (for a detailed review see e.g. \cite{GVY*08})
has lead to the discovery of a close relationship between the multi-component (matrix) NLS equations and the homogeneous
and symmetric spaces~\cite{ForKu*83}. It was shown that the integrable MNLS type models have Lax representation
with the generalized Zakharov--Shabat system as the Lax operator:
\begin{gather}\label{eq:Lax-MNLS}
L\psi (x,t,\lambda ) \equiv i {d\psi \over dx} + (Q(x,t) - \lambda
J)\psi (x,t,\lambda )=0,
\end{gather}
where $J $ is a constant element of the Cartan subalgebra
$\mathfrak{h} \subset \mathfrak{g} $ of the simple Lie algebra
$\mathfrak{g} $ and $Q(x,t) \equiv [J,\widetilde{Q}(x,t)] \in
\mathfrak{g}/\mathfrak{h}$. In other words $Q(x,t) $ belongs to
the co-adjoint orbit $\mathcal{M}_J $ of $\mathfrak{g} $ passing
through $J$. The Hermitian symmetric spaces, compatible with the
NLS dispersion law, are labelled in  Cartan classif\/ication by
${\bf A.III}$, ${\bf C.I}$, ${\bf D.III}$ and ${\bf BD.I}$, see
\cite{Bourb1,Helg}.

In what follows we will assume that the reader is familiar with the theory of simple Lie algebras and
their representations. The choice of $J $ determines the dimension of $\mathcal{M}_J $ which
can be viewed as the phase space of the relevant nonlinear evolution equations (NLEE).  It is equal to the number
of roots of $ \mathfrak{g}$ such that $\alpha (J)\neq 0 $.  Taking into account that if $\alpha  $
is a root, then $-\alpha $ is also a root of $\mathfrak{g} $, so $\dim \mathcal{M}_J $ is always even.

The interpretation of the ISM as a generalized Fourier transforms and the expansion over the so-called
`squared' solutions (see \cite{G} for regular and \cite{VSG*94,G*09} for non-regular $J$) are based
on the spectral theory for Lax operators in the form (\ref{eq:Lax-MNLS}). This allow one to study all the
fundamental properties of the corresponding NLEE's: i)~the description of the class of NLEE related to a~given Lax operator
$L(\lambda)$ and solvable by the ISM; ii)~derivation of the inf\/inite family of integrals of motion; and iii)~their hierarchy
of Hamiltonian structures.

Recently, it has been shown, that some of these MNLS models describe the dynamics of spinor Bose--Einstein condensates in
one-dimensional approximation \cite{IMW04,uiw07}. It also allows an exact description of the dynamics and interaction of bright
solitons with spin degrees of freedom \cite{LLMML05}. Matter-wave solitons are expected to be useful in atom laser, atom interferometry and
coherent atom transport (see e.g.~\cite{UK} and the references therein). Furthermore, a geometric interpretation of the
MNLS models describing spinor Bose--Einstein condensates are given in~\cite{ForKu*83}; Darboux transformation for this special
integrable case is developed in~\cite{LLMML05}.

Along with multi-component NLS-type of systems, generalizations for other hierarchies has also attracted
the interest of the scientif\/ic community: The scalar~\cite{Wad}  and multi-component modif\/ied Korteweg--de Vries hierarchies
over symmetric spaces \cite{AF*87} have been further studied in~\cite{ForKu*83,AF*87,82}.

It is well known that the Lax representation of the MNLS equations takes the form:
\begin{gather*}
[L,M]=0
\end{gather*}
with conveniently chosen operator $M$, see equation~(\ref{eq:3.2}) below. In other words the Lax representation is of pure
Lie algebraic form and therefore the form of the MNLS is independent on the choice of the representation of the relevant
Lie algebra $\mathfrak{g} $. That is why in applying the inverse scattering method (ISM) until now in solving the
direct and the inverse scattering problems for~$L$~(\ref{eq:Lax-MNLS}) only the typical (lowest dimensional exact)
representation of  $\mathfrak{g} $ was used.

Our aim in this paper is to explore the spectral theory of the Lax operator $L$ (\ref{eq:Lax-MNLS}) for dif\/ferent
fundamental representations of the underlying simple Lie algebra $\mathfrak{g}$. We will see that the construction of
such important  for the scattering theory objects like  the fundamental analytic solutions (FAS) depend crucially on the
choice of the representation. This ref\/lects on the formulation of the corresponding Riemann--Hilbert problem (RHP)
and especially on the structure of the so-called dressing factors which allow one to construct the soliton solutions
of the MNLS equations. In turn the dressing factors determine the structure of the singularities of the resolvent of~$L$.
In other words one f\/inds the multiplicities of the discrete eigenvalues of $L$ and the structure of the corresponding
eigensubspaces.

We will pay special attention to the adjoint representation, which
gives the expansion over the so-called `squared' solutions. For
MNLS  related to the orthogonal simple Lie algebras of~${\bf B}_r$
and~${\bf D}_r$ type, we will outline the spectral properties of
the Lax operator in the spinor representations. Another important
tool is the construction of the minimal sets of scattering data~$\mathcal{T}_i$, $i=1,2$, each of which determines uniquely both
the scattering matrix $T(\lambda)$ and the potential~$Q(x)$. Our
remark is that the def\/inition of $\mathcal{T}_i$ used in
\cite{G,VSG*94,Varna04,manev04} is invariant with respect to the
choice of the representation.

The paper is organized as follows: In Section~\ref{section2}, we give some preliminaries about MNLS type equations over symmetric spaces
of {\bf BD.I}-type. Here we summarize the well known facts about their Lax representations, Jost solutions and the scattering
matrix $T(\lambda)$ for the typical representation, see  \cite{G,TMF98}. Next we outline the construction of the FAS and
the relevant RHP which they satisfy. They are constructed by using the Gauss decomposition factors of the scattering matrix
$T(\lambda)$. All our constructions are applied for the class of potentials $Q(x)$ that vanish fast enough for $x\to\pm\infty$.
We f\/inish this section by brief exposition of the simplest type of dressing factors. We also introduce the minimal sets
of scattering data as the minimal sets of coef\/f\/icients $\mathcal{T}_i$, $i=1,2$, which determine the Gauss factors of $T(\lambda)$.
These coef\/f\/icient are representation independent and therefore $\mathcal{T}_i$ determine $T(\lambda)$ and the
corresponding potential~$Q(x)$ in any representation of $\mathfrak{g} $.
In Section~\ref{section3} we describe the spectral properties of the Lax operator in the typical representation for ${\bf BD.I}$ symmetric
spaces and the ef\/fect of the dressing  on the scattering data. We introduce also the kernel of the resolvent $R^\pm(x,y,\lambda)$ of $L$
and use it to derive the completeness relation for the FAS in the typical representation. This relation may also be understood as the
spectral decomposition of $L$.
In Section~\ref{section4} we describe the spectral decomposition of the Lax operator in the adjoint representation: the expansions
over the `squared' solutions and the generating (recursion) operator. Most of the results here have also been known for
some time \cite{AKNS,G,GVY*08}. In particular the coef\/f\/icients of $\mathcal{T}_i$ appear as expansion coef\/f\/icients of the potential
$Q(x)$ over the `squared solutions'. This important fact allows one to treat the ISM as a generalized Fourier transform.
In the next Section~\ref{section5} we study the spectral properties of the same Lax operator in the spinor representation:
starting with the algebraic structure of the Gauss factors for the scattering matrix, associated to $L(\lambda)$, the dressing
factors, etc. Note that the FAS for {\bf BD.I}-type symmetric spaces  are much easier to construct in the spinor representation.
One can view $L_{\rm sp}$ as Lax operator related to the algebra $so(2^r)$ with additional deep reduction that picks up
the spinor representation  of $\mathfrak{g}\simeq so(2r+1) $. In the last Section~\ref{section6} we extend our  results to
a non-fundamental representation.  In order to avoid unnecessary complications we do it on the  example of the $112$-dimensional
representation of~$so(7)$ with highest weight $\omega=3\omega_3 = \frac{3}{2} (e_1+e_2+e_3)$.
The matrix realizations of both adjoint and spinor representations for ${\bf B}_2\simeq so(5, \bbbc)$ and ${\bf B}_7\simeq so(7, \bbbc)$ are presented in Appendices~\ref{appendixA},~\ref{appendixB} and~\ref{appendixC}.

\section{Preliminaries}\label{section2}

We start with some basic facts about the MNLS type models, related to ${\bf BD.I}$ symmetric spaces. Here we will present all result in the typical representation of the corresponding Lie algebra $\mathfrak{g} \simeq {\bf B}_r, {\bf B}_r$. These models admit a Lax representation in the form:
\begin{gather}\label{eq:LM}
L\psi (x,t,\lambda )  \equiv   i\partial_x\psi + (Q(x,t) - \lambda
J)\psi  (x,t,\lambda )=0,\\
 \label{eq:3.2} M\psi (x,t,\lambda )
 \equiv   i\partial_t\psi + (V_0(x,t) +
\lambda V_1(x,t) - \lambda ^2 J)\psi  (x,t,\lambda )=0, \\
V_1(x,t) =  Q(x,t), \qquad V_0(x,t) = i \,{\rm ad}_J^{-1} \frac{d Q}{dx}
+ \frac{1}{2} \left[{\rm ad}_J^{-1} Q, Q(x,t) \right].\nonumber
\end{gather}
Here, in general, $J$ is an element of the corresponding Cartan subalgebra $\mathfrak{h} $ and $Q(x,t)$ is an of\/f-diagonal matrix, taking values in a simple Lie algebra $\mathfrak{g} $. In what follows we assume that $Q(x)\in \mathcal{M}_J$ is a smooth potential  vanishing fast enough
for $x\to\pm\infty$.

Before proceeding further on,  we will f\/ix here the notations and the normalization conditions for the
Cartan--Weyl basis $\{h_k, E_\alpha \}$ of $\mathfrak{g} $
($r=\mbox{rank}\,\mathfrak{g}$) with a root system $\Delta$. We
introduce $h_k\in \mathfrak{h} $, $k=1,\dots,r $ as the Cartan elements
dual to the orthonormal basis $\{e_k\}$ in the root space~$\bbbe^r $
and the Weyl generators $E_\alpha $, $\alpha \in \Delta $. Their
commutation relations are given by \cite{Bourb1,Helg}:
\begin{gather*}
  [h_k,E_\alpha ] = (\alpha ,e_k) E_\alpha , \qquad [E_\alpha
,E_{-\alpha }]={2 \over (\alpha ,\alpha ) } \sum_{k=1}^{r} (\alpha
,e_k) h_k , \nonumber\\
  [E_\alpha ,E_\beta ] = \left\{ \begin{array}{ll} N_{\alpha
,\beta } E_{\alpha +\beta } \quad & \mbox{for} \  \alpha +\beta
\in \Delta, \\ 0 & \mbox{for} \  \alpha +\beta \not\in \Delta
\cup\{0\}. \end{array} \right.
\end{gather*}
Here $\vec{a}=\sum_{k=1}^{r}a_k e_k $ is a  $r $-dimensional
vector dual to $J\in \mathfrak{h} $ and  $(\cdot ,\cdot) $ is the
scalar product in~$\bbbe^r $. The normalization of
the basis is determined by:
\begin{gather*}
  E_{-\alpha } =E_\alpha ^T, \qquad \langle E_{-\alpha },E_\alpha\rangle
={2 \over (\alpha ,\alpha ) }, \qquad
 N_{-\alpha ,-\beta } = -N_{\alpha ,\beta },
\end{gather*}
where $N_{\alpha,\beta }= \pm (p+1) $ and the integer $p\geq 0 $ is such
that $\alpha +s\beta \in\Delta $ for all $s=1,\dots,p $, $ \alpha
+(p+1)\beta \not\in\Delta $ and $\langle \cdot, \cdot \rangle $ is the
Killing form of $\mathfrak{g}$, see \cite{Bourb1,Helg}. The root system
$\Delta $ of $\mathfrak{g} $ is invariant with respect to the group
$W_{\mathfrak{g}} $ of Weyl ref\/lections $S_\alpha $,
\begin{gather*}
S_\alpha \vec{y} = \vec{y} - {2(\alpha ,\vec{y}) \over (\alpha
,\alpha )} \alpha , \qquad \alpha \in \Delta .
\end{gather*}
As it was already mentioned in the Introduction the MNLS equations
correspond to Lax opera\-tor~(\ref{eq:Lax-MNLS}) with non-regular
(constant) Cartan elements $J\in \mathfrak{h}$.  If $J$ is a
regular element of the Cartan subalgebra of $\mathfrak{g}$ then
$\ad_J$ has as many dif\/ferent eigenvalues as is the number of the
roots of the algebra and they are given by $a_j=\alpha_j(J)$,
$\alpha_j\in \Delta$. Such $J $'s can be used to introduce
ordering in the root system by assuming that $\alpha >0 $ if
$\alpha (J)>0 $. In what follows we will assume that all roots for
which $\alpha (J)>0 $ are positive. Obviously one can consider the eigensubspaces of ${\rm ad}_J $ as
grading of the algebra $\mathfrak{g} $.

In the case of symmetric spaces, the corresponding Cartan involution \cite{Helg} provides a grading in $\mathfrak{g} $: $\mathfrak{g} =\mathfrak{g} _0\oplus \mathfrak{g} _1$, where $\mathfrak{g} _0$ is the subalgebra of all elements of $\mathfrak{g} $ commuting with $J$. It contains the Cartan subalgebra $\mathfrak{h} $ and the subalgebra of $\mathfrak{g} \backslash \mathfrak{h} $ spanned on those root subspaces $\mathfrak{g} _\alpha$, such that $\alpha(J)=0$. The set of all such roots is denoted by $\Delta_0$. The corresponding symmetric space is spanned by all root subspaces in $\mathfrak{g} \backslash \mathfrak{g} _0$. Note that one can always use a gauge transformation commuting with $J $ to
remove all components of the potential $Q (x,t)$ that belong to $\mathfrak{ g}_0 $.

For symmetric spaces of ${\bf BD.I}$ type, the potential has the form:
\begin{gather*}
Q=\left(\begin{array}{ccc}  0 & \vec{q}^{T} & 0 \\
  \vec{p} & 0 & s_{0}\vec{q} \\  0 & \vec{p}^{T}s_{0} & 0 \\
\end{array}\right),\qquad J=\mbox{diag}(1,0,\ldots, 0, -1).
\end{gather*}
For $n=2r+1$ the $n$-component vectors  $\vec{q}$ and $\vec{p}$ have the form
$ \vec{q} = (q_1,\dots , q_r,q_0,q_{\bar{r}},\dots ,  q_{\bar{1}})^T$, $ \vec{p} =
(p_1,\dots , p_r,p_0,p_{\bar{r}},\dots ,  p_{\bar{1}})^T$,
while the matrix $s_0 =S_0^{(n)}$ enters in the def\/inition of
$so(n)$:
$ X\in so(n)$, $X + S_0^{(n)} X^T S_0^{(n)} =0$, where
\begin{gather*}
S_0^{(n)} =  \sum_{s=1}^{n+1} (-1)^{s+1} E_{s, n+1-s}^{(n)}.
\end{gather*}
For $n=2r$ the $n$-component vectors  $\vec{q}$ and $\vec{p}$ have the form
$ \vec{q} = (q_1,\dots , q_r,q_{\bar{r}},\dots ,  q_{\bar{1}})^T$, $ \vec{p} =
(p_1,\dots , p_r,p_{\bar{r}},\dots ,  p_{\bar{1}})^T$ and
\begin{gather*}
S_0^{(n )} = \sum_{s=1}^{r} (-1)^{s+1} \big(E_{s, n+1-s}^{(n)}+
E_{n+1-s,s}^{(n)}\big).
\end{gather*}
With this def\/inition of orthogonality the Cartan
subalgebra generators are represented by diago\-nal matrices. By
$E^{(n)}_{s,p}$ above we mean $n\times n$ matrix whose matrix
elements are $(E^{(n)}_{s,p})_{ij}=\delta_{si}\delta_{pj}$.

Let us comment brief\/ly on the algebraic structure of the Lax pair,
which is related to the symmetric space $SO(n+2)/(SO(n)\times
SO(2))$. The element of the Cartan subalgebra $J$, which is dual
to $e_1 \in \bbbe^r$ allows us to introduce a grading in it:
$\mathfrak{g}=\mathfrak{g}_0\oplus \mathfrak{g}_1$ which
satisf\/ies:
\begin{gather*}
[X_1, X_2]\in \mathfrak{g}_0, \qquad [X_1,Y_1]\in \mathfrak{g}_1,
\qquad [Y_1,Y_2]\in \mathfrak{g}_0,
\end{gather*}
for any choice of the elements $X_1, X_2\in \mathfrak{g}_0$ and
$Y_1, Y_2 \in \mathfrak{g}_1$. The grading splits the set of
positive roots of $so(n)$ into two subsets $\Delta^+= \Delta_0^+
\cup \Delta_1^+$ where $\Delta_0^+$ contains all the positive
roots of $\mathfrak{g}$ which are orthogonal to $e_1$, i.e.\
$(\alpha,e_1)=0$; the roots in $\beta\in\Delta_1^+$ satisfy
$(\beta,e_1)=1$. For more details see the appendix  below and~\cite{Helg}.

The Lax pair can be considered in any representation of $so(n)$, then the potential
$Q$ will take the form:
\begin{gather*}
Q(x,t) =\sum_{\alpha\in\Delta_1^+} \left( q_{\alpha}(x,t)
E_{\alpha} + p_{\alpha}(x,t) E_{-\alpha} \right).
\end{gather*}
Next we introduce $n$-component `vectors' formed by the Weyl
generators of $so(n+2)$ corresponding to the roots in
$\Delta_1^+$:
\begin{gather*}
\vec{E}_1^\pm = ( E_{\pm (e_1-e_2)}, \dots , E_{\pm (e_1-e_r)},
E_{\pm e_1}, E_{\pm(e_1+e_r)}, \dots , E_{\pm (e_1+e_2)} ),
\end{gather*}
for $n=2r+1$ and
\begin{gather*}
\vec{E}_1^\pm = ( E_{\pm (e_1-e_2)}, \dots , E_{\pm (e_1-e_r)},
 E_{\pm(e_1+e_r)}, \dots , E_{\pm (e_1+e_2)} ),
\end{gather*}
for $n=2r$. Then the generic form of the potentials $Q(x,t)$
related to these type of symmetric  spaces can be written as sum
of two ``scalar'' products
\begin{gather*}
Q(x,t) = (\vec{q}(x,t) \cdot \vec{E}_1^+) + (\vec{p}(x,t) \cdot
\vec{E}_1^-) .
\end{gather*}
In terms of these notations the generic MNLS type equations
connected to ${\bf BD.I}$ acquire the form
\begin{gather}
i \vec{q}_t  + \vec{q}_{xx} + 2 (\vec{q},\vec{p}) \vec{q} -
 (\vec{q},s_0\vec{q}) s_0\vec{p} =0,\nonumber \\
i \vec{p}_t  - \vec{p}_{xx} - 2 (\vec{q},\vec{p}) \vec{p} +
(\vec{p},s_0\vec{p}) s_0\vec{q} =0.\label{eq:4.2}
\end{gather}
The Hamiltonian for the MNLS equations (\ref{eq:4.2}) with the canonical reduction $\vec{p}=\epsilon \vec{q}^{*}$, $\epsilon=\pm
1$ imposed, is given by:
\begin{gather*}
H_{{\rm MNLS}}=\int_{-\infty}^\infty d x\left((\partial_{x}\vec{q},\partial_{x}\vec{q^{*}})-\epsilon
(\vec{q},\vec{q^{*}})^2+ \frac{\epsilon}{2} (\vec{q},s_0\vec{q})(\vec{q^{*}},s_{0}\vec{q^{*}})\right).
\end{gather*}

\subsection[Direct scattering problem for $L$]{Direct scattering problem for $\boldsymbol{L}$}\label{section2.1}

The starting point for solving the direct and the inverse scattering problem (ISP) for
$L$ are  the so-called Jost solutions,  which are def\/ined by their asymptotics (see, e.g.~\cite{VSG2}
and the references therein):
\begin{gather}\label{eq:Jost}
\lim_{x \to -\infty} \phi(x,t,\lambda) e^{  i \lambda J x }=\openone, \qquad  \lim_{x \to \infty}\psi(x,t,\lambda) e^{  i
\lambda J x } = \openone
 \end{gather}
and the scattering matrix $T(\lambda)$ is def\/ined by $T(\lambda,t)\equiv
\psi^{-1}\phi(x,t,\lambda)$. Here we assume that the potential $q(x,t)$ is tending to zero fast enough, when $|x|\to \infty$. The special choice of $J$ results in the
fact that the Jost solutions and the scattering matrix take values
in the corresponding orthogonal Lie group $SO(n+2)$. One can use the following block-matrix
structure of $T(\lambda,t)$
\begin{gather*}
T(\lambda,t) = \left( \begin{array}{ccc} m_1^+ & -\vec{b}^-{}^T & c_1^- \\
\vec{b}^+ & {\bf T}_{22} & - s_0\vec{B}^- \\ c_1^+ & \vec{B}^+{}^Ts_0 & m_1^- \\
\end{array}\right), \qquad \hat{T}(\lambda,t) = \left( \begin{array}{ccc} m_1^- & \vec{B}^-{}^T & c_1^- \\
-\vec{B}^+ & {\bf \hat{T}}_{22} &  s_0\vec{b}^- \\ c_1^+ & -\vec{b}^+{}^Ts_0 & m_1^+ \\
\end{array}\right),
\end{gather*}
where $\vec{b}^\pm (\lambda,t)$ and $\vec{B}^\pm (\lambda,t)$ are $n$-component vectors, ${\bf T}_{22}(\lambda)$ and
${\bf \hat{T}}_{22} (\lambda)$ are $n \times n$ block matrices, and $m_1^\pm
(\lambda)$, $c_1^\pm (\lambda)$ are scalar functions. Here  and below by `hat' we will denote taking the inverse, i.e.\
$\hat{T}(\lambda, t) =T^{-1}(\lambda)$. Such parametrization is compatible with the generalized Gauss
decompositions \cite{Helg} of $T(\lambda,t)$.

With this notations we introduce the generalized Gauss factors of
$T(\lambda)$ as follows:
\begin{gather*}
T(\lambda,t) = T^-_J D^+_J \hat{S}^+_J  = T^+_J D^-_J \hat{S}^-_J ,
\\
T^-_J = e^{\left(\vec{\rho}^+ , \vec{E}^- \right)}   =
\left( \begin{array}{ccc}  1 & 0 & 0 \\ \vec{\rho}^+ & \openone & 0 \\
c_1^{'',+} & \vec{\rho}^{+,T}s_0 & 1 \\ \end{array} \right),  \qquad
T^+_J =e^{\left(-\vec{\rho}^- , \vec{E}^+ \right)}   = \left(
\begin{array}{ccc}  1 & -\vec{\rho}^{-,T} & c_1^{'',-} \\ 0
& \openone & -s_0\vec{\rho}^- \\ 0 & 0 & 1 \\ \end{array} \right),\nonumber\\
 S^+_J = e^{\left(\vec{\tau}^+ , \vec{E}^+ \right)}  = \left(
\begin{array}{ccc}  1 & \vec{\tau}^{+,T} & c_1^{',-} \\ 0 & \openone & s_0\vec{\tau}^+
\\ 0 & 0 & 1 \\ \end{array} \right),   \qquad S^-_J = e^{\left(-\vec{\tau}^- , \vec{E}^- \right)}  =
\left( \begin{array}{ccc}  1 & 0 & 0 \\ -\vec{\tau}^- & \openone & 0 \\
c_1^{',+} & -\vec{\tau}^{-,T}s_0 & 1 \\ \end{array} \right),
\\
D_J^+  = \left( \begin{array}{ccc} m_1^+ & 0 & 0 \\ 0 & \m_2^+ & 0 \\
0 & 0 & 1/m_1^+ \end{array} \right),  \qquad  D_J^-  =
\left( \begin{array}{ccc} 1/m_1^- & 0 & 0 \\ 0 &  \m_2^- & 0 \\
0 & 0 & m_1^- \end{array} \right),\nonumber\\
c_1^{'',\pm}  = \frac{1}{2}(\vec{\rho}^{\pm,T} s_0
\vec{\rho}^\pm) ,  \qquad c_1^{',\pm}  =
\frac{1}{2}(\vec{\tau}^{\mp,T} s_0 \vec{\tau}^\mp),\nonumber
\end{gather*}
where
\begin{gather}
c_1^- = \frac{m_1^-}{2} (\vec{\rho}^{-,T} s_0 \vec{\rho}^-)  =
\frac{m_1^+}{2} (\vec{\tau}^{+,T} s_0 \vec{\tau}^+) , \qquad c_1^+
= \frac{m_1^+}{2} (\vec{\rho}^{+,T} s_0 \vec{\rho}^+) =
\frac{m_1^-}{2} (\vec{\tau}^{-,T} s_0 \vec{\tau}^-) ,\nonumber\\
\vec{\rho}^- =\frac{\vec{B}^-}{m_1^-}, \qquad \vec{\tau}^-
=\frac{\vec{B}^+}{m_1^-}, \qquad \vec{\rho}^+ =\frac{\vec{b}^+}{m_1^+}, \qquad \vec{\tau}^+
=\frac{\vec{b}^-}{m_1^+},\nonumber\\
\m_2^+ = {\bf T}_{22} + \frac{\vec{b}^+ \vec{b}^-{}^T  }{m_1^+},
\qquad \m_2^- = {\bf T}_{22} +
 \frac{s_0\vec{b}^- \vec{b}^+{}^T s_0 }{m_1^-}.\label{eq:25.1a}
\end{gather}
These notations satisfy a number of  relations which ensure that both $T(\lambda)$ and its inverse $\hat{T}(\lambda)$ belong to the
corresponding orthogonal group $SO(n+2)$ and that $T(\lambda)\hat{T}(\lambda) =\openone$. Some of them take the form:
\begin{alignat*}{3}
& m_1^+m_1^- + (\vec{b}^-, \vec{B}^+) + c_1^+ c_1^- =1, \qquad && \vec{b}^+ \vec{B}^-{}^T + {\bf T}_{22} s_0 {\bf T}_{22}^T s_0 +
s_0 \vec{B}^- \vec{b}^+{}^Ts_0 = \openone,&\nonumber \\
& 2m_1^+ c_1^- - \vec{b}^{-,T} s_0 \vec{b}^-=0 , \qquad && 2m_1^- c_1^+ - \vec{B}^{+,T} s_0 \vec{B}^+ =0,&\nonumber \\
& m_1^-\vec{b}^+ - {\bf T}_{22} \vec{B}^+ - s_0 \vec{B}^- c_1^+ =0 , \qquad && m_1^+\vec{B}^- - {\bf \hat{T}}_{22}^T \vec{b}^- - s_0
\vec{b}^+ c_1^- =0.& 
\end{alignat*}
Important tools for reducing the ISP to a Riemann--Hilbert problem (RHP) are the fundamental analytic solution (FAS) $\chi^{\pm}
(x,t,\lambda )$. Their construction is based on the generalized Gauss decomposition of $T(\lambda,t)$, see \cite{ZMNP,G,TMF98}:
\begin{gather}\label{eq:FAS_J}
\chi ^\pm(x,t,\lambda)= \phi (x,t,\lambda) S_{J}^{\pm}(t,\lambda )
= \psi (x,t,\lambda ) T_{J}^{\mp}(t,\lambda ) D_J^\pm (\lambda).
\end{gather}
More precisely, this construction ensures that $\xi^\pm(x,\lambda)=\chi^\pm(x,\lambda) e^{i\lambda Jx}$ are
analytic functions of $\lambda$ for $\lambda \in \bbbc_\pm$. Here $S_{J}^{\pm} $, $T_{J}^{\pm} $ upper- and lower-block-triangular
matrices, while $D_J^\pm(\lambda)$ are block-diagonal matrices with the same block structure as $T(\lambda,t)$ above. Skipping
the details, we give here the explicit expressions of the Gauss factors in terms of the matrix elements of $T(\lambda,t)$
\begin{gather*}
 S_J^\pm (t,\lambda )= \exp \big( \pm (\vec{\tau}^\pm
(\lambda,t) \cdot \vec{E}_1^\pm ) \big), \qquad  T_J^\pm
(t,\lambda )= \exp \big( \mp (\vec{\rho}^\mp (\lambda,t) \cdot
\vec{E}_1^\pm ) \big), \nonumber
\end{gather*}
where
\begin{gather*}
\vec{\tau}^+ (\lambda ,t) = \frac{\vec{b}^-}{m_1^+}, \qquad
\vec{\tau}^- (\lambda ,t) = \frac{\vec{B}^+}{m_1^-}, \qquad
\vec{\rho}^+ (\lambda ,t) = \frac{\vec{b}^+}{m_1^+}, \qquad
\vec{\rho}^-(\lambda ,t) = \frac{\vec{B}^-}{m_1^-},
\end{gather*}
and
\begin{alignat*}{3}
& {\bf T}_{22} =\m_2^+ - \frac{\vec{b}^+ \vec{b}^-{}^T }{2m_1^+}, \qquad && {\bf T}_{22}  =\m_2^- -  \frac{s_0\vec{b}^- \vec{b}^+{}^T
s_0 }{2m_1^-}, & \nonumber\\
 & {\bf \hat{T}}_{22}=\hat{\m}_2^+ - \frac{s_0\vec{b}^- \vec{b}^+{}^T s_0}{2m_1^+}, \qquad& &  {\bf \hat{T}}_{22}
 =\hat{\m}_2^- - \frac{\vec{B}^+ \vec{B}^-{}^T }{2m_1^-}. & 
\end{alignat*}
The two analyticity regions $\bbbc_+$ and $\bbbc_-$ are separated by the real line. The continuous spectrum of $L$ f\/ills
in the real line and has multiplicity $2$, see Section~\ref{section3.3} below.

If $Q(x,t) $ evolves according to (\ref{eq:4.2}) then the scattering matrix and its elements satisfy the following linear
evolution equations
\begin{gather*}
 i\frac{d\vec{b}^{\pm}}{d t} \pm \lambda ^2 \vec{b}^{\pm}(t,\lambda ) =0, \qquad i\frac{d\vec{B}^{\pm}}{d t}
\pm \lambda ^2\vec{B}^{\pm}(t,\lambda ) =0, \qquad
 i\frac{d m_1^{\pm}}{d t}  =0, \qquad  i \frac{d \m_2^{\pm}}{d t}  =0,
\end{gather*}
so the block-diagonal matrices $D^{\pm}(\lambda)$ can be considered as generating functionals of the integrals of motion.
It is well known \cite{DrSok,BeSat,ForKu*83,G} that generic nonlinear evolution equations related to a simple Lie
algebra $\mathfrak{g} $ of rank $r$ possess $r$ series of integrals of motion in involution; thus for them one
can prove complete integrability~\cite{BeSat}. In our case we can consider as generating functionals of
integrals of motion  all $(2r-1)^2$ matrix elements of $\m_2^\pm(\lambda)$, as well as $m_1^\pm(\lambda)$ for $\lambda \in \bbbc_\pm$.
However they can not be all in involution. Such situation is characteristic for the superintegrable models. It is due to
the degeneracy of the dispersion law of~(\ref{eq:4.2}). We remind that $D^\pm_J(\lambda)$ allow analytic
extension for $\lambda\in \bbbc_\pm$ and that their zeroes and poles determine the discrete eigenvalues of $L$.

\subsection[Riemann-Hilbert problem and minimal set of scattering data for $L$]{Riemann--Hilbert problem and minimal set of scattering data for $\boldsymbol{L}$}\label{section2.2}

The FAS for real $\lambda$ are linearly related
\begin{gather}\label{eq:rhp0}
\chi^+(x,t,\lambda) =\chi^-(x,t,\lambda) G_{0,J}(\lambda,t),
\qquad G_{0,J}(\lambda,t)=\hat{S}^-_J(\lambda,t)S^+_J(\lambda,t) .
\end{gather}
One can rewrite equation~(\ref{eq:rhp0}) in an equivalent form for the
FAS $\xi^\pm(x,t,\lambda)=\chi^\pm (x,t,\lambda)e^{i\lambda Jx }$
which satisfy the equation:
\begin{gather}\label{eq:xi}
i\frac{d\xi^\pm}{dx} + Q(x)\xi^\pm(x,\lambda) -\lambda [J,
\xi^\pm(x,\lambda)]=0, \qquad \lim_{\lambda \to \infty} \xi^\pm(x,t,\lambda) = \openone.
\end{gather}
Then these FAS satisfy
\begin{gather}\label{eq:rhp1}
\xi^+(x,t,\lambda) =\xi^-(x,t,\lambda) G_J(x,\lambda,t), \qquad
G_{J}(x,\lambda,t) =e^{-i\lambda
Jx}G^-_{0,J}(\lambda,t)e^{i\lambda Jx} .
\end{gather}
Obviously the sewing function $G_J(x,\lambda,t)$ is uniquely
determined by the Gauss factors $S_J^\pm (\lambda,t)$. Equation (\ref{eq:rhp1}) is a
Riemann--Hilbert problem (RHP) in multiplicative form. Since the Lax operator has no discrete
eigenvalues,  $\det\xi^\pm(x,\lambda)$ have no zeroes for
$\lambda\in \bbbc_\pm$ and $\xi^\pm(x,\lambda)$ are regular
solutions of the RHP. As it is well known the regular solution
$\xi^\pm(x,\lambda)$ of the RHP is uniquely determined.

Given the solutions $\xi^\pm(x,t,\lambda)$ one recovers $Q(x,t)$
via the formula
\begin{gather*}
Q(x,t) = \lim_{\lambda\to\infty} \lambda \big( J - \xi^\pm J
\widehat{\xi}^\pm(x,t,\lambda)\big) = [J,\xi_1(x)],
\end{gather*}
which is obtained from equation (\ref{eq:xi}) taking the limit
$\lambda\to\infty$. By $\xi_1(x)$ above we have denoted $\xi_1(x)
= \lim\limits_{\lambda\to\infty}\lambda(\xi(x,\lambda)-\openone)$.

If the potential $Q(x,t)$ is such that the Lax operator $L$ has
no discrete eigenvalues, then the minimal set of scattering data
is given by one of the sets $\mathfrak{T}_i$, $i=1,2$
\begin{gather}
  \mathfrak{T}_1 \equiv \{ \rho_\alpha^+(\lambda,t),
\rho_\alpha^-(\lambda,t), \  \alpha\in\Delta_1^+, \  \lambda
\in \bbbr\},  \nonumber \\
 \mathfrak{T}_2 \equiv \{
\tau_\alpha^+(\lambda,t), \tau_\alpha^-(\lambda,t), \
\alpha\in\Delta_1^+, \  \lambda \in \bbbr\}.  \label{eq:T_i}
\end{gather}
Any of these sets determines uniquely the scattering matrix $T(\lambda,t)$ and
the corresponding potential $Q(x,t)$. For more details, we refer to \cite{VSG2,FaTa,ZMNP} and references therein.

Most of the known examples of MNLS on symmetric spaces are obtained after imposing  the  reduction:
\begin{gather*}
 Q(x,t)=K_0^{-1} Q^\dag(x,t)K_0 , \qquad K_0=\exp \left( \frac{\pi
i}{2}\sum_{j=1}^r (3+\epsilon_j)H_j \right),\nonumber \\
 p_k=\tilde{K}_0 q_k^*, \qquad \tilde{K}_0 =\diag (K_{0,22}, \dots
, K_{0,n+1,n+1})  
\end{gather*}
or in components $p_k=\epsilon_1\epsilon_k q_k^*$. As a
consequence the corresponding MNLS takes the form:
\begin{gather*}
i \vec{q}_t + \vec{q}_{xx} + 2 (\vec{q},\tilde{K}_0\vec{q}^*)
\vec{q} -  (\vec{q},s_0\vec{q}) s_0\tilde{K}_0\vec{q}^* =0.
\end{gather*}
The scattering data are restricted by $\vec{\rho}^-(\lambda,t)=
\tilde{K}_0 \vec{\rho}^{+,*}(\lambda,t)$ and
$\vec{\tau}^-(\lambda,t)= \tilde{K}_0\vec{\tau}^{+,*}(\lambda,t)$.

If all $\epsilon_j=1$ then the reduction becomes the ``canonical'' one:
$Q(x,t)=Q^\dag (x,t)$ and $\vec{\rho}^-(\lambda,t)=
\vec{\rho}^{+,*}(\lambda,t)$ and $\vec{\tau}^-(\lambda,t)=
\vec{\tau}^{+,*}(\lambda,t)$.

\subsection{Dressing factors and soliton solutions}\label{section2.3}

The main goal of the dressing method \cite{ZaSha,vgrn,I04,GGK05a,GGK05b,PLA126} is, starting from a known
solutions $\chi^\pm_0(x,t,\lambda)$ of $L_0(\lambda) $ with
potential $Q_{(0)}(x,t)$ to construct new singular solutions
$\chi^\pm_1(x,t,\lambda )$ of $L$ with a potential $Q_{(1)}(x,t)$
with two additional singularities located at prescribed positions
$\lambda _1^\pm $; the reduction $\vec{p} =\vec{q}^{*}$ ensures
that $\lambda_1^-=(\lambda_1^+)^*$. It is related to the regular
one by a dressing factor $u(x,t,\lambda )$
\begin{gather}\label{eq:Dressfactor}
\chi^{\pm}_1(x,t,\lambda)=u(x,\lambda) \chi^{\pm}_0(x,t,\lambda)
u_{-}^{-1}(\lambda ), \qquad u_-(\lambda )=\lim_{x\to -\infty }
u(x,\lambda ).
\end{gather}
Note that $u_-(\lambda )$ is a block-diagonal matrix. The dressing
factor $u(x,\lambda ) $ must satisfy the equation
\begin{gather}\label{eq:u-eq}
i\partial_x u + Q_{(1)}(x) u - u Q_{(0)}(x)- \lambda
[J,u(x,\lambda)] =0,
\end{gather}
and the normalization condition $\lim\limits_{\lambda \to\infty }
u(x,\lambda ) =\openone $. The construction of $u(x,\lambda )\in
SO(n+2) $ is based on an appropriate anzatz specifying explicitly
the form of its $\lambda $-dependence (see \cite{Za*Mi,GGK05b} and the
references therein). Here we will consider a special choice of dressing factors:
\begin{gather}
u(x,\lambda)=\openone+(c(\lambda)-1)P (x)
+\left(\frac{1}{c(\lambda)}-1\right)
 \overline{P}(x), \nonumber\\
\overline{P} = S_0^{-1}P^TS_0, \qquad c(\lambda)={\lambda-\lambda_1^+\over \lambda - \lambda_1^-},
\label{eq:rank1}
\end{gather}
where  $P(x)$ and $\overline{P}(x)$ are mutually orthogonal
projectors with rank~1. More specif\/ically we have
\begin{gather}\label{eq:31}
 P(x,t)= {|n(x,t)\rangle  \langle m(x,t)|\over \langle n(x,t)|
m(x,t)\rangle},
\end{gather}
where $\langle m(x,t)|=\langle m_0|(\chi_0^-(x,\lambda_1^-))^{-1}$ and
$|n(x,t)\rangle=\chi_0^+(x,\lambda_1^+)|n_0\rangle$; $\langle m_0|$ and
$|n_0\rangle$ are (constant) polarization vectors \cite{vgrn}.
Taking the limit $\lambda \to \infty$ in equation (\ref{eq:u-eq}) we
get that
\[ Q_{(1)}(x,t) - Q_{(0)}(x,t) = (\lambda_1^- - \lambda_1^+) [ J,
P(x,t)- \overline{P}(x,t)] .\] Below  we assume that $Q_{(0)}=0$
and impose $\bbbz_2$ reduction condition
\begin{gather*}
KQK^{-1}=Q,\qquad
K=\diag(\epsilon_1,\epsilon_2,\dots, \epsilon_{r-1},1,\epsilon_r,\epsilon_{r-1},\dots, \epsilon_2,\epsilon_1).
\end{gather*}
This in its turn leads to $\lambda_1^\pm = \mu \pm i\nu$ and
$|m_a\rangle=K_a|n_a\rangle^*$, $a=1,\dots, n+2$. The polarization vectors $\langle m(x,t)|$ and $|n(x,t)\rangle$ are parameterised as follows:
\[
\langle m(x,t)|=(m_1(x,t),m_2(x,t),\dots, m_r(x,t),0,m_r(x,t), \dots, m_2(x,t), m_1(x,t))
\]
and
\[
|n(x,t)\rangle =(n_1(x,t),n_2(x,t),\dots, n_r(x,t),0,n_r(x,t),  \dots, n_2(x,t), n_1(x,t))^T.
\]
As a result one gets:
\begin{gather*}
q_k^{(\rm 1s)}(x,t) = - 2i\nu \big( P_{1k}(x,t) + (-1)^k
P_{\bar{k},n+2} (x,t) \big),
\end{gather*}
where $\bar{k}=n+3-k$. For more details on the soliton solutions satisfying  the standard reduction $Q(x,t)=Q^\dag (x,t)$ see
\cite{gkv08,I04,kagg07,GGK05a,GGK05b,G*09}.

The ef\/fect of the dressing on the scattering data (\ref{eq:25.1a}) is as follows:
\begin{alignat*}{3}
& \vec{\rho}_1^+ ={\vec{b}^+\over m_1^+}={1\over c(\lambda)}\vec{\rho}_0^+, \qquad && \vec{\rho}_1^- ={\vec{B}^-\over m_1^-}= c(\lambda)\vec{\rho}_0^-;&\nonumber\\
& \vec{\tau}_1^+ ={\vec{b}^-\over m_1^+}={1\over c(\lambda)}\vec{\tau}_0^+, \qquad&& \vec{\tau}_1^- ={\vec{B}^+\over m_1^-}= c(\lambda)\vec{\tau}_0^-.&
\end{alignat*}
 Applying $N$ times the dressing method, one gets a Lax operator with $N$ pairs of prescribed discrete eigenvalues $\lambda_j^\pm$, $j=1,\dots,N$. The minimal sets of scattering data for the ``dressed'' Lax operator contains in addition the  discrete eigenvalues $\lambda_j^\pm$, $j=1,\dots,N$ and the corresponding ref\/lection/transmission coef\/f\/icient at these points:
\begin{gather}
 \mathfrak{T}^\prime_1 \equiv \{ \rho_\alpha^+(\lambda,t),
\rho_\alpha^-(\lambda,t),\rho_{\alpha,j}^+(t), \rho_{\alpha,j}^+(t), \lambda_j^\pm, \ \alpha\in\Delta_1^+, \  \lambda
\in \bbbr, \  j=1, \dots,N\}, \nonumber \\   \mathfrak{T}^\prime_2 \equiv \{
\tau_\alpha^+(\lambda,t), \tau_\alpha^-(\lambda,t), \tau_{\alpha,j}^+(t), \tau_{\alpha,j}^+(t), \lambda_j^\pm, \ \alpha\in\Delta_1^+, \ \lambda
\in \bbbr, \  j=1, \dots,N\}. \label{eq:T_i-dress}
\end{gather}

\section[Resolvent and spectral decompositions in the
typical representation of $\mathfrak{g}\simeq B_r$]{Resolvent and spectral decompositions\\ in the
typical representation of $\boldsymbol{\mathfrak{g}\simeq B_r}$}\label{section3}

Here we f\/irst formulate the interrelation between the `naked' and dressed FAS.
We will use these relations to determine the order of pole singularities of the
resolvent which of course will inf\/luence the contribution of the discrete spectrum
to the completeness relation.

We f\/irst start with the simplest and more general case of the generalized Zakharov--Shabat
system in which all eigenvalues of $J$ are dif\/ferent. Then we discuss the additional
construction necessary to treat cases when $J$ has vanishing and/or equal eigenvalues.

\subsection{The ef\/fect of dressing on the scattering data}\label{section3.1}

Let us f\/irst determine the ef\/fect of dressing on the Jost solutions with the simplest dressing
factor $u_1(x,\lambda)$. In what follows below we denote the Jost solutions corresponding to the
regular solutions of RHP by $\psi_0(x,\lambda)$
and the dressed one by  $\psi_1(x,\lambda)$. In order to preserve the def\/inition in equation (\ref{eq:Jost}) we put:
\begin{gather*}
\psi_1(x,\lambda)= u_1(x,\lambda) \psi_0(x,\lambda) \hat{u}_{1,+}(\lambda), \qquad
\phi_1(x,\lambda)= u_1(x,\lambda) \phi_0(x,\lambda) \hat{u}_{1,-}(\lambda),
\end{gather*}
where $u_{1,\pm}(x,\lambda)=\lim\limits_{x\to\pm\infty} u_1(x,\lambda)$. We will also use the fact that
$u_{1,\pm}(\lambda)$ are $x$-independent elements belonging to the Cartan subgroup of $\mathfrak{g} $.
For the typical representation of $so(2r+1)$ and for the case in which only two singularities $\lambda_1^\pm$
are added we have:
\begin{gather}
u(x,\lambda)= \exp \left( \ln c_1(\lambda)(P_1 - \bar{P}_1)\right), \nonumber\\
u_{1,+}(\lambda) = \openone + (c_1(\lambda)-1) E_{11} + \left( \frac{1}{c_1(\lambda)} -1 \right) E_{n+2,n+2}, \nonumber\\
u_{1,-}(\lambda) = \openone + (c_1(\lambda)-1) E_{n+2,n+2} + \left( \frac{1}{c_1(\lambda)} -1 \right) E_{11}, \nonumber\\
u_{1,\pm}(\lambda) =\exp \left( \pm\ln c_1(\lambda) J\right).\label{eq:upm1}
\end{gather}
Then from equation (\ref{eq:FAS_J}) we get:
\begin{gather*}
\chi_1^\pm(x,\lambda)  = u_1(x,\lambda)\chi^\pm_0 (x,\lambda)\hat{u}_{1,-}(\lambda).
\end{gather*}
As a consequence we f\/ind that
\begin{alignat}{3}
& T_1(\lambda) = u_{1,+}(\lambda) T_0(\lambda)  \hat{u}_{1,-}(\lambda) , \qquad&&
D_1^\pm(\lambda) = u_{1,+}(\lambda) D_0(\lambda)  \hat{u}_{1,-}(\lambda) ,& \nonumber\\
& S_1^\pm(\lambda) = u_{1,-}(\lambda) S_0(\lambda)  \hat{u}_{1,-}(\lambda) , \qquad&&
T_1^\pm(\lambda) = u_{1,+}(\lambda) T_0(\lambda)  \hat{u}_{1,+}(\lambda) .&\label{eq:T0T1}
\end{alignat}
One can repeat  the dressing procedure $N$ times by using the dressing factor:
\begin{gather}\label{eq:un}
u(x,\lambda) = u_N(x,\lambda)  u_{N-1}(x,\lambda) \cdots  u_1(x,\lambda) .
\end{gather}
Note that  the projector $P_k$ of the $k$-th dressing factor has the form of
(\ref{eq:31}) but the $x$-dependence of the polarization vectors is determined
by the $k-1$ dressed FAS:
\begin{gather}
\chi_{k}^\pm (x,\lambda)   = u_k(x,\lambda) u_{k-1}(x,\lambda)\cdots  u_{1}(x,\lambda)
\chi_0^\pm (x,\lambda) \hat{u}_{1,-}(\lambda) \cdots \hat{u}_{k-1,-}(\lambda) \hat{u}_{k,-} (\lambda), \nonumber\\
u_k(x,\lambda)  =\openone+(c_k(\lambda)-1)P_k (x)+\left(\frac{1}{c_k(\lambda)}-1\right)
\overline{P}_k(x),   \qquad \overline{P}_k = S_0^{-1}P_k^TS_0,\label{eq:chiu}
\\
c_k(\lambda)={\lambda-\lambda_k^+\over \lambda_k - \lambda_1^-}, \qquad
P_k(x,t) = {|n_k(x,t)\rangle  \langle m_k(x,t)|\over \langle n_k(x,t)| m_k(x,t)\rangle}, \nonumber\\
\langle m_k(x,t)|  =\langle m_{0,k}|(\chi_{k-1}^-(x,\lambda_k^-))^{-1},  \qquad
|n_k(x,t)\rangle  =\chi_{k-1}^+(x,\lambda_k^+)|n_{0,k}\rangle,\label{eq:cPk}
\end{gather}
and $\langle m_{0,k}|$ and $|n_{0,k}\rangle$ are (constant) polarization vectors

Using equation (\ref{eq:chiu}) and assuming that all $\lambda_j \in \bbbc_\pm$ are dif\/ferent  we can treat the general
case  of  RHP with $2N$ singular points.  The corresponding relations between the `naked' and dressed
FAS are:
\begin{alignat}{3}
& T(\lambda) = u_-(\lambda) T_0(\lambda)  \hat{u}_{-}(\lambda) , \qquad&&
D^\pm(\lambda) = u_{+}(\lambda) D_0(\lambda)  \hat{u}_{-}(\lambda) ,&\nonumber\\
& S^\pm(\lambda) = u_{-}(\lambda) S_0(\lambda)  \hat{u}_{-}(\lambda) , \qquad&&
T^\pm(\lambda) = u_{+}(\lambda) T_0(\lambda)  \hat{u}_{+}(\lambda) ,&\label{eq:T0TN}
\end{alignat}
where
\begin{gather}
u_{+}(\lambda) = \openone + (c(\lambda)-1) E_{1,1} + \left( \frac{1}{c(\lambda)} -1 \right) E_{n+2,n+2}, \nonumber\\
u_{-}(\lambda) = \openone + (c(\lambda)-1) E_{n+2,n+2} + \left( \frac{1}{c(\lambda)} -1 \right) E_{1,1}, \nonumber\\
u_{\pm}(\lambda) =\exp \left( \pm\ln c(\lambda) J\right), \qquad c(\lambda) =\prod_{k=1}^N c_k(\lambda).\label{eq:upm}
\end{gather}
In components equations (\ref{eq:T0TN}) give:
\begin{alignat*}{3}
& m_1^+(\lambda) = m_{1,0}^+(\lambda) c^2(\lambda), \qquad && m_1^-(\lambda) = \frac{m_{1,0}^-(\lambda)}{ c^2(\lambda)}, &\nonumber\\
& \rho_1^+(\lambda) = \frac{\rho_{0}^+(\lambda)}{ c(\lambda)}, \qquad && \rho_1^-(\lambda)  =  c(\lambda)\rho_{0}^-(\lambda), &\nonumber\\
& \tau_1^+(\lambda) = \frac{\tau_{0}^+(\lambda)}{ c(\lambda)}, \qquad && \tau_1^-(\lambda) =  c(\lambda)\tau_{0}^-(\lambda),& 
\end{alignat*}
and $\m_2^\pm(\lambda)=\m_{2,0}^\pm(\lambda)$.

In what follows we will need the residues of $u(x,\lambda)\chi_0^\pm(x,\lambda)$ and its inverse $\hat{\chi}_0^\pm(x,\lambda)\hat{u}(x,\lambda)$
at $\lambda=\lambda_k^\pm$ respectively.
From equations (\ref{eq:chiu}) and (\ref{eq:cPk}) we get:
\begin{gather}
u(x,\lambda)\chi_0^+(x,\lambda)  \simeq \frac{(\lambda_k^- - \lambda_k^+) \chi^{+,(k)}(x) }{\lambda - \lambda_k^+} +
\dot{\chi}^{+,(k)}(x) + \mathcal{O}(\lambda - \lambda_k^+), \nonumber\\
u(x,\lambda)\chi_0^-(x,\lambda)  \simeq \frac{(\lambda_k^+ - \lambda_k^-) \chi^{-,(k)}(x) }{\lambda - \lambda_k^-} +
\dot{\chi}^{-,(k)}(x) + \mathcal{O}(\lambda - \lambda_k^-), \nonumber\\
\hat{\chi}_0^+(x,\lambda)\hat{u}(x,\lambda) \simeq \frac{(\lambda_k^- - \lambda_k^+) \hat{\chi}^{+,(k)}(x) }{\lambda - \lambda_k^+} +
\widehat{\dot{\chi}}^{+,(k)}(x) + \mathcal{O}(\lambda - \lambda_k^+), \nonumber\\
\hat{\chi}_0^-(x,\lambda)\hat{u}(x,\lambda) \simeq \frac{(\lambda_k^+ - \lambda_k^-) \hat{\chi}^{-,(k)}(x) }{\lambda - \lambda_k^-} +
\widehat{\dot{\chi}}^{-,(k)}(x) + \mathcal{O}(\lambda - \lambda_k^-),\label{eq:chi-res}
\end{gather}
where
\begin{gather}
\chi^{+,(k)}(x)  = u_N(x,\lambda_k^+)\cdots u_{k+1}(x,\lambda_k^+) \bar{P}_k \chi^+_{(k-1)} (x,\lambda_k^+), \nonumber\\
\chi^{-,(k)}(x)  = u_N(x,\lambda_k^-)\cdots u_{k+1}(x,\lambda_k^-) P_k \chi^-_{(k-1)} (x,\lambda_k^-), \nonumber\\
\hat{\chi}^{+,(k)}(x)  = \hat{\chi}^+_{(k-1)} (x,\lambda_k^+) \hat{u}_{k+1}(x,\lambda_k^+)\cdots \hat{u}_{N}(x,\lambda_k^+) P_k,\nonumber \\
\hat{\chi}^{-,(k)}(x)  = \hat{\chi}^-_{(k-1)} (x,\lambda_k^-) \hat{u}_{k+1}(x,\lambda_k^-)\cdots \hat{u}_{N}(x,\lambda_k^-) \bar{P}_k.\label{eq:chi-k}
\end{gather}
These results will be used below to f\/ind the residues of the resolvent at $\lambda=\lambda_k^\pm$.

\subsection[Spectral decompositions for the generalized Zakharov-Shabat system:
$sl(n)$-case]{Spectral decompositions for the generalized Zakharov--Shabat system:\\
$\boldsymbol{sl(n)}$-case}\label{section3.2}

The FAS are the basic tool in constructing the spectral theory of the corresponding
Lax operator. For the generic Lax operators related to the $sl(n)$ algebras:
\begin{gather*}
L_{\rm gen}\equiv i\frac{\partial \chi_{\rm gen} }{\partial x} + (Q_{\rm gen}(x) -
\lambda J_{\rm gen})\chi_{\rm gen}(x,\lambda) =0, \qquad (Q_{\rm gen})_{jj}(x)=0 ,
\end{gather*}
this theory is well developed, see \cite{ZMNP,LMP,G,G-cm}. In the generic case
all eigenvalues of $J_{\rm gen} =\diag (J_1,J_2,\dots,J_{n+2})$ are dif\/ferent
and non-vanishing:
\begin{gather*}
J_1 > J_2 > \cdots > J_k>0 > J_{k+1} > \cdots > J_{n+2}, \qquad \tr J_{\rm gen} =0.
\end{gather*}
The Jost solutions $\psi_{\rm gen}(x,\lambda)$,  $\phi_{\rm gen}(x,\lambda)$, the scattering matrix  $T_{\rm gen}(\lambda)$
and the FAS $\chi_{\rm gen}^\pm (x,\lambda)$ are introduced by \cite{Sha,ZMNP} (see also \cite{Sh,G,G-cm,VSG2,LMP,G*09}):
\begin{alignat*}{3}
&\lim_{x\to-\infty}\phi_{\rm gen}(x,\lambda) e^{iJ_{\rm gen}x\lambda} =\openone, \qquad &&
\lim_{x\to\infty}\psi_{\rm gen}(x,\lambda) e^{iJ_{\rm gen}x\lambda} =\openone, & \nonumber\\
&T_{\rm gen} (\lambda) =  \psi_{\rm gen}^{-1}(x,\lambda) \phi_{\rm gen}(x,\lambda), &&&\nonumber \\
&\chi_{\rm gen}^\pm (x,\lambda)  = \phi_{\rm gen}(x,\lambda) S_{\rm gen}^\pm (\lambda), \qquad &&
\chi_{\rm gen}^\pm (x,\lambda)  = \psi_{\rm gen}(x,\lambda) T_{\rm gen}^\mp (\lambda)D_{\rm gen}^\pm (\lambda), &
\end{alignat*}
where $S_{\rm gen}^\pm (\lambda)$,  $T_{\rm gen}^\pm (\lambda)$ and  $D_{\rm gen}^\pm (\lambda)$ are the factors in the
Gauss decompositions of  $T_{\rm gen} (\lambda)$:
\begin{gather*}
T_{\rm gen} (\lambda) = T_{\rm gen}^-(\lambda) D_{\rm gen}^+ (\lambda) \hat{S}_{\rm gen}^+ (\lambda)
 = T_{\rm gen}^+ (\lambda) D_{\rm gen}^- (\lambda) \hat{S}_{\rm gen}^- (\lambda).
\end{gather*}
More specif\/ically $S_{\rm gen}^+ (\lambda)$ and $T_{\rm gen}^+ (\lambda)$ (resp.\
$S_{\rm gen}^- (\lambda)$ and $T_{\rm gen}^- (\lambda)$) are upper (resp.\ lower) trian\-gular  matrices
whose diagonal elements are equal to~1. The diagonal matrices  $D_{\rm gen}^+ (\lambda) $ and $D_{\rm gen}^- (\lambda) $ allow
analytic extension in the upper and lower half planes respectively.

The dressing factors $u_k(x,\lambda)$ are the simplest possible ones \cite{ZMNP}:
\begin{gather*}
u_{k,\rm gen}(x,\lambda) = \openone + (c_k(\lambda) -1 )P_k(x), \qquad u^{-1}_{k,\rm gen}(x,\lambda) = \openone +
\left( \frac{1}{c_k(\lambda)} -1 \right) P_k(x),
\end{gather*}
where the rank-1 projectors $P_k(x)$ are expressed through the regular solutions analogously to equations (\ref{eq:un})--(\ref{eq:cPk})
with $\chi_0^\pm(x,\lambda_k^\pm)$ replaced by $\chi_{0,\rm gen}^\pm(x,\lambda_k^\pm)$

The relations between the dressed and `naked' scattering data are the same like in equation~(\ref{eq:T0T1}) only now
the asymptotic values $u_{\rm gen; \pm}(\lambda)$ are dif\/ferent. Assuming that all projectors~$P_k$ have rank 1 we get:
\begin{gather*}
u_{\rm gen; \pm}(\lambda)  = \prod_{k=1}^N u_{k, \rm gen; \pm}(\lambda),  \qquad
u_{k, \rm gen; \pm}(\lambda)  = \openone + (c_k(\lambda) -1 )P_{k, \pm}, \nonumber\\
P_{k,+}  =  E_{s_k, s_k},  \qquad P_{k,-}  =E_{p_k, p_k},
\end{gather*}
where $s_k$ (resp.~$p_k$) labels the  position of the f\/irst (resp. the last) non-vanishing component of the
polarization vector $|n_{0,k}\rangle $.

Using the FAS we introduce the resolvent $R_{\rm gen}(\lambda ) $ of $L_{\rm gen}$ in the form:
\begin{gather*}
R_{\rm gen}(\lambda ) f(x) = \int_{-\infty }^{\infty } R_{\rm gen}(x,y,\lambda ) f(y).
\end{gather*}
The kernel $R_{\rm gen}(x,y,\lambda ) $ of the resolvent is given by:
\begin{gather*}
R_{\rm gen}(x,y,\lambda ) = \left\{ \begin{array}{ll}
R_{\rm gen}^+(x,y,\lambda ) \ & \mbox{for} \ \lambda \in \bbbc^+, \\
R_{\rm gen}^-(x,y,\lambda ) \ & \mbox{for} \  \lambda \in \bbbc^-,
\end{array} \right.
\end{gather*}
where
\begin{gather*}
R_{\rm gen}^\pm (x,y,\lambda ) = \pm i\chi_{\rm gen} ^\pm(x,\lambda ) \Theta ^\pm (x-y)
\hat{\chi }_{\rm gen}^\pm (y,\lambda ),\nonumber \\
\Theta ^\pm(z)  =  \theta (\mp z) \Pi_0 - \theta (\pm z) (\openone -\Pi_0),
\qquad \Pi_0 = \sum_{s=1}^k E_{ss}, 
\end{gather*}

\begin{theorem}\label{th:R2}
Let $Q(x) $ be a potential of $L$ which falls off fast enough for $x\to\pm\infty$
and the corresponding RHP has a finite number of simple singularities
at the  points $\lambda _j^\pm \in \bbbc_\pm$, i.e.\
$\chi_{\rm gen}^\pm(x,\lambda)$ have  simple poles and zeroes at  $\lambda _j^\pm $. Then

\begin{enumerate}\itemsep=0pt

\item[$1)$] $R_{\rm gen}^\pm (x,y,\lambda ) $ is an analytic function of $\lambda  $ for
$\lambda \in \bbbc_\pm $ having pole singularities at $\lambda _j^\pm \in
\bbbc_\pm $;

\item[$2)$] $R_{\rm gen}^\pm (x,y,\lambda ) $ is a kernel of a bounded integral operator
for $\im \lambda \neq 0 $;

\item[$3)$] $R_{\rm gen} (x,y,\lambda ) $ is an uniformly bounded function for $\lambda
\in\bbbr $ and provides the kernel of an unbounded integral operator;

\item[$4)$] $R_{\rm gen}^\pm (x,y,\lambda ) $ satisfy the equation:
\begin{gather*}
L_{\rm gen}(\lambda ) R_{\rm gen}^\pm (x,y,\lambda )=\mbox{\rm $\openone$} \delta (x-y).
\end{gather*}
\end{enumerate}
\end{theorem}

Skipping the details (see \cite{G-cm}) we will formulate below
the completeness relation for the eigenfunctions of the
Lax operator $L_{\rm gen}$. It is derived by applying the contour  integration method
(see e.g.~\cite{GeKu,AKNS}) to the integral:
\begin{gather*}
\mathcal{J}_{\rm gen}(x,y)={1\over 2\pi i}\oint_{\gamma _+} d\lambda
R_{\rm gen}^{+}(x,y,\lambda)- {1\over 2\pi i}\oint_{\gamma _-} d\lambda
R_{\rm gen}^{-}(x,y,\lambda),
\end{gather*}
where the contours $\gamma _\pm$ are shown on the Fig.~\ref{fig:1} and
has the form:

\begin{figure}[t]
\centerline{\includegraphics{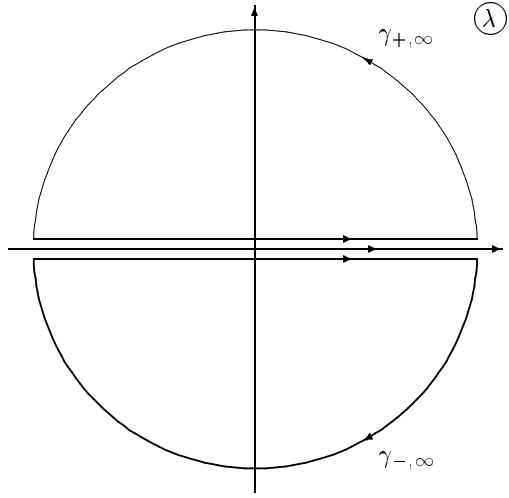}}

\caption{The contours $\gamma _\pm =\bbbr\cup\gamma_{\pm\infty }$.}
\label{fig:1}
\end{figure}

The explicit form of the dressing factors $u_{\rm gen}(x,\lambda )$ makes it obvious that
the kernel of the resolvent has only simple poles at $\lambda =\lambda_k^\pm$. Therefore
the f\/inal form of the completeness relation for the Jost solutions of $L_{\rm gen}$ takes the form
\cite{G-cm}:
\begin{gather}
\delta(x-y)  \sum_{s=1}^{n} {1  \over a_s } E_{ss} \nonumber\\
 \qquad{} = {1\over  2\pi}\int_{-\infty}^\infty d\lambda\left\{ \sum_{s=1}^{k_0}
|\chi_{\rm gen}^{[s]+}(x,\lambda)\rangle \langle \hat\chi_{\rm gen}^{[s]+}(y,\lambda)|
 - \sum_{s=k_0+1}^{n} |\chi_{\rm gen}^{[s]-}(x,\lambda)\rangle \langle
\hat\chi_{\rm gen}^{[s]-}(y,\lambda)|\right\} \nonumber\\
\qquad {}+  \sum_{j=1}^N
\left( \Res_{\lambda=\lambda_j^+}R^+(x,y) +  \Res_{\lambda=\lambda_j^-}R^-(x,y) \right).\label{eq:3.44}
\end{gather}
It is easy to check that the residues in (\ref{eq:3.44}) can be expressed by the
properly normalized eigenfunctions of $L_{\rm gen}$ corresponding to the eigenvalues  $\lambda_j^\pm$
\cite{G-cm}.

Thus we conclude that the continuous spectrum of $L_{\rm gen}$ has multiplicity $n$ and f\/ills up the whole
real axis $\bbbr$ of the complex $\lambda$-plane; the discrete eigenvalues of $L_{\rm gen}$ constructed using
the dressing factors $u_{\rm gen}(x,\lambda)$ are simple and the resolvent kernel $R_{\rm gen}(x,y,\lambda)$
has poles of order one at $\lambda=\lambda_k^\pm$.

\subsection{Resolvent and spectral decompositions for BD.I-type Lax
operators}\label{section3.3}

In our case $J$ has $n$ vanishing eigenvalues which makes the problem
more dif\/f\/icult.

We can rewrite the Lax operator in the form:
\begin{gather*}
i \frac{\partial \chi_1 }{\partial x} + \vec{q}^T \vec{\chi}_0  = \lambda \chi_1, \nonumber\\
i \frac{\partial \vec{\chi}_0 }{\partial x} + \vec{q}\, ^* \chi_1 +s_0\vec{q} \chi_{-1}  = 0, \nonumber\\
i \frac{\partial \chi_{-1} }{\partial x} + \vec{q}\, ^\dag s_0 \vec{\chi}_0  = \lambda \chi_{-1}, 
\end{gather*}
where we have split the eigenfunction $\chi(x,\lambda)$ of $L$ into three according to the
natural block-matrix structure compatible with $J$: $\chi(x,\lambda)= \left(\chi_1, \vec{\chi}_0^T, \chi_{-1}
\right)^T $. Note that the  equation for $\vec{\chi}_0$ can not be treated as eigenvalue equations;
they can be formally integrated with:
\begin{gather*}
\vec{\chi}_0(x,\lambda ) = \vec{\chi}_{0,\rm as} + i \int_{}^{x} dy  \left( \vec{q}\, ^* \chi_1 +s_0\vec{q} \chi_{-1}
\right),
\end{gather*}
which eventually casts the Lax operator into
the following integro-dif\/ferential system with non-degenerate $\lambda$ dependence:
\begin{gather*}
i \frac{\partial \chi_1 }{\partial x} + i \vec{q}^T (x) \int_{}^{x} dy  \left( \vec{q}\, ^* \chi_1 +s_0\vec{q}
\chi_{-1} \right) (y,\lambda)  = \lambda \chi_1,\nonumber \\
i \frac{\partial \chi_{-1} }{\partial x} + i \vec{q}\,^\dag (x) s_0  \int_{}^{x} dy  \left( \vec{q}\, ^* \chi_1 +s_0\vec{q}
\chi_{-1} \right) (y,\lambda) = -\lambda \chi_{-1}.
\end{gather*}
Similarly we can treat the operator which is adjoint to $L$ whose FAS $\hat{\chi}(x,\lambda)$ are
the inverse to $\chi(x,\lambda)$, i.e.\ $\hat{\chi}(x,\lambda) = \chi^{-1}(x,\lambda)$. Splitting
each of the rows of $\hat{\chi}(x,\lambda)$ into components as follows $\hat{\chi}(x,\lambda)=
(\hat{\chi}_1, \hat{\vec{\chi}}_0, \hat{\chi}_{-1})$ we get:
\begin{gather*}
i \frac{\partial \hat{\chi}_1 }{\partial x} - (\hat{\vec{\chi}}_0, \vec{q}\,^*)  - \lambda \hat{\chi}_1 =0, \nonumber\\
i \frac{\partial \hat{\vec{\chi}}_0 }{\partial x} -\hat{\chi}_1 \vec{q}^T - \hat{\chi}_{-1} \vec{q}\,^\dag s_0 = 0, \nonumber\\
i \frac{\partial \hat{\chi}_{-1} }{\partial x} -(\hat{\vec{\chi}}_0,s_0 \vec{q}) -\lambda \hat{\chi}_{-1} =0.
\end{gather*}
Again the  equation for $\hat{\vec{\chi}}_0$ can   be formally  integrated with:
\begin{gather*}
\hat{\vec{\chi}}_0(x,\lambda ) = \hat{\vec{\chi}}_{0,\rm as} + i \int_{}^{x} dy  \left( \hat{\chi}_1(y,\lambda) \vec{q} ^T(y) +
 \hat{\chi}_{-1}(y,\lambda) \vec{q}\, ^\dag (y) s_0 \right).
\end{gather*}
Now we get the following integro-dif\/ferential system with non-degenerate $\lambda$ dependence
\begin{gather*}
i \frac{\partial \hat{\chi}_1 }{\partial x} - i  \int_{}^{x} dy  \left( \hat{\chi}_1(y,\lambda) (\vec{q}^T (y),\vec{q}\,^* (x))
+\hat{\chi}_{-1}(y,\lambda) (\vec{q}\,^\dag (y) s_0\vec{q}\,^* (x))\right) + \lambda \hat{\chi}_1  =0, \nonumber\\
i \frac{\partial \hat{\chi}_{-1} }{\partial x} - i  \int_{}^{x} dy  \left( \hat{\chi}_1(y,\lambda) (\vec{q}^T (y)s_0\vec{q} (x))
+\hat{\chi}_{-1}(y,\lambda) (\vec{q}\,^\dag (y) ,\vec{q} (x))\right) - \lambda \hat{\chi}_{-1}  =0.
\end{gather*}
Now we are ready to generalize the standard approach to the case of BD.I-type Lax operators.

The kernel $R (x,y,\lambda ) $ of the resolvent is given by:
\begin{gather*}
R (x,y,\lambda ) = \left\{ \begin{array}{ll}
R^+(x,y,\lambda ) \ & \mbox{for}  \ \lambda \in \bbbc^+, \\
R^-(x,y,\lambda ) \ & \mbox{for} \ \lambda \in \bbbc^-,
\end{array} \right.
\end{gather*}
where
\begin{gather*}
R^\pm (x,y,\lambda ) = \pm i\chi ^\pm(x,\lambda ) \Theta ^\pm (x-y) \hat{\chi }^\pm (y,\lambda ),  \nonumber\\
\Theta ^\pm(z)  =  \theta (\mp z) E_{11} - \theta (\pm z) (\openone -E_{11}).
\end{gather*}
The completeness relation for the eigenfunctions of the
Lax operator $L $  is derived by applying the contour  integration method
(see e.g.~\cite{GeKu,AKNS}) to the integral:
\begin{gather*}
\mathcal{J}'(x,y)={1\over 2\pi i}\oint_{\gamma _+} d\lambda\,
\Pi_1 R^{+}(x,y,\lambda)- {1\over 2\pi i}\oint_{\gamma _-} d\lambda\,
\Pi_1 R^{-}(x,y,\lambda),
\end{gather*}
where the contours $\gamma _\pm$ are shown on the Fig.~\ref{fig:1} and $\Pi_1 = E_{11}+E_{n+2,n+2}$.
Using equations~(\ref{eq:chi-res}) and~(\ref{eq:chi-k}) we are able to check that the kernel of the resolvent
has poles of second order at $\lambda=\lambda_k^\pm$. Therefore the completeness relation takes the form:
\begin{gather*}
\Pi_1 \delta(x-y)   = {1\over  2\pi}\int_{-\infty}^\infty d\lambda \Pi_1 \left\{
|\chi_{\rm gen}^{[1]+}(x,\lambda)\rangle \langle \hat\chi_{\rm gen}^{[1]+}(y,\lambda)|
 -  |\chi_{\rm gen}^{[n+2]-}(x,\lambda)\rangle \langle
\hat\chi_{\rm gen}^{[n+2]-}(y,\lambda)|\right\} \nonumber\\
\phantom{\Pi_1 \delta(x-y)   =}{}  + 2i \sum_{j=1}^N
\left\{ \Res_{\lambda=\lambda_k^+} R^+(x,y,\lambda) +  \Res_{\lambda=\lambda_k^-} R^-(x,y,\lambda)  \right\},
\end{gather*}
where
\begin{gather*}
 \Res_{\lambda=\lambda_k^\pm} R^\pm(x,y,\lambda) =\pm (\lambda_k^- - \lambda_k^+) \Pi_1 \left(
\chi^{+,(k)}(x) \hat{\dot{\chi}}^{+,(k)}(y) + \dot{\chi}^{+,(k)}(x) \hat{\chi}^{+,(k)}(y) \right).
\end{gather*}
In other words  the continuous spectrum of $L$ has multiplicity $2$ and f\/ills up the whole
real axis~$\bbbr$ of the complex $\lambda$-plane; the discrete eigenvalues of~$L $ constructed using
the dressing factors $u(x,\lambda)$ (\ref{eq:upm}) lead to second order poles of resolvent kernel $R_{\rm gen}(x,y,\lambda)$
at $\lambda=\lambda_k^\pm$.

\section[Resolvent and spectral decompositions in the
adjoint representation of $\mathfrak{g}\simeq B_r$]{Resolvent and spectral decompositions\\ in the
adjoint representation of $\boldsymbol{\mathfrak{g}\simeq B_r}$}\label{section4}

The simplest realization of $L$ in the adjoint representation is to make use of the adjoint action of $Q(x)-\lambda J$ on
$\mathfrak{g}$:
\begin{gather*}
L_{\rm ad} e_{\rm ad} \equiv i \frac{\partial e_{\rm ad} }{\partial x}
+ \left[ Q(x) -\lambda J_{\rm ad} , e_{\rm ad}(x,\lambda)\right] =0.
\end{gather*}
Note that the eigenfunctions of $L_{\rm ad}$ take values in the Lie algebra $\mathfrak{g}$. They are known also as the
`squared solutions' of $L$ and appear in a natural way in the analysis of the transform from the potential $Q(x,t)$ to
the scattering data of $L$, \cite{AKNS}; see also \cite{G,G-cm} and the numerous references therein.

The  idea for the interpretation of the ISM as a generalized Fourier transform  was
launched in \cite{AKNS}. It is based on the Wronskian relations which allow one to maps the potential $Q(x,t)$
onto the minimal sets of scattering data $\mathcal{T}_i$. These ideas have been generalized also to the symmetric spaces,
see \cite{G,Varna04,G*09} and the references therein.

The `squared solutions' that play the role of generalized exponentials are  determined by the FAS and the Cartan--Weyl basis of the
corresponding algebra as follows. First we introduce:
\begin{gather*}
e_{\alpha,\rm ad}^\pm(x,\lambda) = \chi^\pm E_\alpha \hat{\chi}^\pm(x,\lambda) , \qquad
e_{j,\rm ad}^\pm(x,\lambda) = \chi^\pm H_j  \hat{\chi}^\pm(x,\lambda) ,
\end{gather*}
where $\chi^\pm(x,\lambda)$ are the FAS of $L$ and $E_\alpha$, $H_j$ form the Cartan--Weyl basis of $\mathfrak{g}$.
Next we note that in the adjoint representation  $J_{\rm ad} \cdot \equiv {\rm ad}_J\cdot \equiv [J, \cdot] $ has kernel. Just like in the
previous section we have to project out that kernel, i.e.\ we need to introduce the projector:
\begin{gather*}
\pi_J X \equiv {\rm ad}_J^{-1} {\rm ad}_J X,
\end{gather*}
for any $X\in \mathfrak{g} $. In particular, choosing $\mathfrak{g}\simeq so(2r+1) $ and $J$ as in equation (\ref{eq:LM})
we f\/ind that the potential $Q$  provides a generic element of the image of $\pi_J$, i.e.\ $\pi_J Q\equiv Q$.

Next, the analysis of the Wronskian relations allows one to introduce two sets of squared solutions:
\begin{gather*}
\bPsi_\alpha ^\pm = \pi_J (\chi^\pm(x,\lambda) E_\alpha
\hat{\chi}^\pm(x,\lambda)), \qquad \bPhi_\alpha ^\pm =
\pi_J(\chi^\pm(x,\lambda) E_{-\alpha} \hat{\chi}^\pm(x,\lambda)), \qquad
\alpha \in \Delta_1^+.
\end{gather*}
We remind that the set $\Delta_1$ contains all roots of $so(r+1)$ for which $\alpha(J)\neq 0$.

Let us introduce the sets of `squared solutions':
\begin{gather*}
\{\bPsi \} = \{\bPsi \}_{\rm c} \cup \{\bPsi \}_{\rm d}, \qquad
\{\bPhi \} = \{\bPhi \}_{\rm c} \cup \{\bPhi \}_{\rm d},\\
\{\bPsi \}_{\rm c}  \equiv \left\{ \bPsi ^+_{\alpha}(x,\lambda),
\  \bPsi ^-_{-\alpha}(x,\lambda), \  i<r, \  \lambda \in
\bbbr \right\},
\nonumber\\
\{\bPsi \}_{\rm d}  \equiv \left\{\bPsi ^+_{\alpha;j}(x), \
\dot{\bPsi }^+_{\alpha ;j}(x),\   \bPsi ^-_{-\alpha;j}(x),\
\dot{\bPsi }^-_{-\alpha;j}(x)\right\}_{j=1}^{N},
\\
\{\bPhi \}_{\rm c}  \equiv \left\{ \bPhi ^+_{-\alpha}(x,\lambda),
\  \bPhi ^-_{\alpha}(x,\lambda), \  i<r, \  \lambda \in
\bbbr \right\},\nonumber\\
\{\bPhi \}_{\rm d}  \equiv \left\{\bPhi ^+_{-\alpha;j}(x), \
\dot{\bPhi }^+_{-\alpha;j}(x),\   \bPhi ^-_{\alpha;j}(x),\
\dot{\bPhi }^-_{\alpha;j}(x)\right\}_{j=1}^{N},
\end{gather*}
where the subscripts `c' and `d' refer to the continuous and
discrete spectrum of $L $.

Each of the above two sets are complete sets of functions in the space of allowed
potentials. This fact can be  proved by applying the contour integration method to the
integral
\begin{gather*}
\mathcal{J}_G(x,y) = \frac{1}{2\pi i} \oint_{\gamma_+} d\lambda
G^+(x,y,\lambda) - \frac{1}{2\pi i} \oint_{\gamma_-} d\lambda
G^-(x,y,\lambda),
\end{gather*}
where the Green function is def\/ined by:
\begin{gather*}
G^\pm(x,y,\lambda)  = G_1^\pm(x,y,\lambda) \theta(y-x) - G_2^\pm(x,y,\lambda) \theta(x-y), \nonumber\\
G_1^\pm(x,y,\lambda)   = \sum_{\alpha \in \Delta_1^+} \bPsi_{\pm\alpha} ^\pm
(x,\lambda) \otimes \bPhi_{\mp \alpha}^\pm (y,\lambda), \nonumber\\
G_2^\pm (x,y,\lambda)  = \sum_{\alpha \in \Delta_0\cup
\Delta_{1}^{-}} \bPhi_{\pm\alpha}^\pm(x,\lambda)\otimes \bPsi_{\mp\alpha}^\pm (y,\lambda) + \sum_{j=1}^r \h_{j}^\pm
(x,\lambda)\otimes \h_{j}^\pm(y,\lambda), \nonumber\\
\h_{j}^\pm (x,\lambda)  = \chi^\pm(x,\lambda) H_j \hat{\chi}^\pm (x,\lambda).
\end{gather*}
Skipping the details we give the result \cite{Varna04,Porto08}:
\begin{gather}
\delta(x-y)\Pi_{0J} = {1\over \pi} \int_{-\infty}^\infty
d \lambda (G_1^+(x,y,\lambda) - G_1^-(x,y,\lambda) ) \nonumber\\
\phantom{\delta(x-y)\Pi_{0J} =}{} - 2i \sum_{j=1}^{N} (G_{1,j}^+(x,y) + G_{1,j}^-(x,y) ),\label{eq:5.23}
\end{gather}
where
\begin{gather*}
\Pi_{0J} =\sum_{\alpha \in \Delta_{1}^{+}}( E_{\alpha}\otimes
E_{-\alpha} - E_{-\alpha}\otimes E_{\alpha}) ,
\\
G_{1,}j^\pm (x,y) = \sum_{\alpha \in \Delta_{1}^{+}}
(\dot{\bPsi}_{\pm \alpha;j}^\pm (x)\otimes \bPhi_{\mp \alpha;j}^\pm (y)
+  \bPsi_{\pm \alpha;j}^\pm (x)\otimes \dot{\bPhi}_{\mp \alpha;j}^\pm (y)).
\end{gather*}

\subsection{Expansion over the `squared solutions'}\label{section4.1}

The completeness relation of the `squared solutions' allows one to expand any function over
the `squared solutions'

Using the Wronskian relations one can derive the expansions over
the `squared solutions' of two important functions. Skipping the
calculational details  we formulate the results \cite{Varna04}.
The expansion of $Q(x) $ over the systems $ \{\bPhi ^\pm \}$ and $
\{\bPsi ^\pm \}$ takes the form:
\begin{gather}
Q(x)  = \frac{i}{\pi } \int_{-\infty }^{\infty } \rd\lambda
\sum_{\alpha \in \Delta_1^+} \left( \tau^+_{\alpha }(\lambda )
\bPhi_{\alpha } ^+(x, \lambda )
-\tau_{\alpha }^-(\lambda )  \bPhi_{-\alpha } ^-(x, \lambda ) \right) \nonumber\\
\phantom{Q(x)  =}{} + 2\sum_{k=1}^{N} \sum_{\alpha \in \Delta_1^+}
\left(\tau^+_{\alpha ;j} \bPhi_{\alpha ;j} ^+(x) + \tau^-_{\alpha
;j} \bPhi_{-\alpha ;j} ^-(x)\right),\label{eq:49.4}
\\
Q(x) =- {\ri \over \pi } \int_{-\infty }^{\infty } \rd\lambda
\sum_{\alpha \in \Delta_1^+} \left( \rho^+_{\alpha }(\lambda )
\bPsi_{-\alpha } ^+(x, \lambda )
-\rho_{\alpha }^-(\lambda )  \bPsi_{\alpha } ^-(x, \lambda ) \right)\nonumber \\
\phantom{Q(x) =}{} - 2\sum_{k=1}^{N} \sum_{\alpha \in \Delta_1^+}
\left(\rho^+_{\alpha ;j} \bPsi_{-\alpha ;j} ^+(x) + \rho^-_{\alpha
;j} \bPsi_{\alpha ;j} ^-(x)\right).\label{eq:49.5}
\end{gather}
The next  expansion is of ${\rm ad}_J^{-1}\delta Q(x) $ over the
systems $ \{\bPhi ^\pm \}$ and $ \{\bPsi ^\pm \}$:
\begin{gather}
{\rm ad}_J^{-1}\delta Q(x) = {\ri\over 2\pi } \int_{-\infty }^{\infty
} \rd\lambda \sum_{\alpha \in \Delta_1^+} \left(
\delta\tau^+_{\alpha }(\lambda ) \bPhi_{\alpha } ^+(x, \lambda )
+ \delta \tau_{\alpha }^-(\lambda )  \bPhi_{-\alpha } ^-(x, \lambda ) \right)\nonumber \\
\phantom{{\rm ad}_J^{-1}\delta Q(x) =}{} + \sum_{k=1}^{N} \sum_{\alpha \in \Delta_1^+} \left(\delta
W^+_{\alpha ;j}(x) -
\delta'W^-_{-\alpha ;j}(x) \right),\label{eq:50.6} \\
{\rm ad}_J^{-1}\delta Q(x)  = {\ri  \over 2\pi } \int_{-\infty
}^{\infty } \rd\lambda \sum_{\alpha \in \Delta_1^+} \left(\delta
\rho^+_{\alpha }(\lambda ) \bPsi_{-\alpha } ^+(x, \lambda )
+ \delta\rho_{\alpha }^-(\lambda )  \bPsi_{\alpha } ^-(x, \lambda ) \right) \nonumber \\
\phantom{{\rm ad}_J^{-1}\delta Q(x)  =}{} +\sum_{k=1}^{N} \sum_{\alpha \in \Delta_1^+}
\left(\delta\tilde{W}^+_{-\alpha ;j}(x) -
\delta\tilde{W}^-_{\alpha ;j}(x)\right),\label{eq:51.4}
\end{gather}
where
\begin{gather*}
\delta W^\pm_{\pm\alpha ;j}(x)=\delta\lambda_j^\pm
\tau^\pm_{\alpha ;j} \dot{\bPhi}_{\pm\alpha;j} ^\pm(x) + \delta
\tau^\pm_{\alpha ;j} \bPhi_{\pm\alpha;j} ^\pm(x), \\
\delta \tilde{W}^\pm_{\mp \alpha ;j}(x)=\delta\lambda_j^\pm
\rho^\pm_{\alpha;j} \dot{\bPsi}_{\mp\alpha;j} ^\pm (x) + \delta
\rho^\pm_{\alpha;j} \bPsi_{\mp\alpha;j} ^\pm(x)
\end{gather*}
and $\bPhi_{\pm\alpha;j} ^\pm(x)=\bPhi_{\pm\alpha}
^\pm(x,\lambda_j^\pm)$, $\dot{\bPhi}_{\pm\alpha;j} ^\pm(x)
=\partial_\lambda \bPhi_{\pm\alpha} ^\pm(x,\lambda)
|_{\lambda=\lambda_j^\pm} $.

The expansions (\ref{eq:49.4}), (\ref{eq:49.5}) is another way to
establish the one-to-one correspondence bet\-ween~$Q(x) $ and each
of the minimal sets of scattering data $\mathcal{T}_1 $ and
$\mathcal{T}_2 $ (\ref{eq:T_i}). Likewise the expansions~(\ref{eq:50.6}),~(\ref{eq:51.4}) establish the one-to-one
correspondence between the variation of the potential $\delta Q(x)
$ and the variations of the scattering data $\delta \mathcal{T}_1
$ and $\delta \mathcal{T}_2 $.

The expansions (\ref{eq:50.6}), (\ref{eq:51.4}) have a special
particular case when one considers the class of variations of
$Q(x,t)$ due to the evolution in $t$. Then
\begin{gather*}
\delta Q(x,t)\equiv Q(x,t+\delta t)-Q(x,t) = \frac{\partial
Q}{\partial t}\delta t + \mathcal{O}\big((\delta t)^2\big).
\end{gather*}
Assuming that $\delta t$ is small and keeping only the f\/irst order
terms in $\delta t$ we get the expansions for ${\rm ad}_J^{-1}Q_t$.
They are obtained from (\ref{eq:50.6}), (\ref{eq:51.4}) by
replacing $\delta \rho_\alpha^\pm (\lambda)$ and $\delta
\tau_\alpha^\pm (\lambda)$ by $\partial_t \rho_\alpha^\pm
(\lambda)$ and $
\partial_t \rho_\alpha^\pm (\lambda)$.

\subsection{The generating operators}\label{section4.2}

To complete the analogy between the standard Fourier transform and
the expansions over the `squared solutions' we need the analogs of
the operator $D_0=-\ri \rd/\rd x $. The operator $D_0 $ is the one
for which $\re^{\ri\lambda x} $ is an eigenfunction:  $D_0
\re^{\ri\lambda x}=\lambda \re^{\ri\lambda x} $. Therefore it is
natural to introduce the generating operators $\Lambda _\pm $
through:
\begin{alignat*}{4}
& (\Lambda _+-\lambda )\bPsi_{-\alpha}^{+} (x,\lambda ) = 0, \qquad &&
(\Lambda _+-\lambda )\bPsi_{\alpha}^{-} (x,\lambda ) = 0, \qquad &&
(\Lambda _+ -\lambda_j^\pm )\bPsi_{\mp\alpha;j}^{+}(x)= 0, & \nonumber\\
& (\Lambda _--\lambda )\bPhi_{\alpha}^{+} (x,\lambda ) = 0, \qquad &&
(\Lambda _--\lambda )\bPhi_{-\alpha}^{-} (x,\lambda ) = 0, \qquad &&
(\Lambda _+ -\lambda_j^\pm )\bPhi_{\pm\alpha;j}^{+}(x)= 0, & 
\end{alignat*}
where the generating operators $\Lambda _\pm $ are given by:
\begin{gather*}
\Lambda _\pm X(x) \equiv {\rm ad}_{J}^{-1} \left( \ri {\rd X \over \rd
x} + \ri \left[ Q(x), \int_{\pm\infty }^{x} \rd y\, [Q(y),
X(y)]\right] \right).
\end{gather*}
The rest of the squared solutions are not eigenfunctions of
neither $\Lambda _+ $ nor $\Lambda _-$:
\begin{alignat*}{3}
& (\Lambda _+ -\lambda_j^+ )
\dot{\bPsi}_{-\alpha;j}^{+}(x)=\bPsi_{-\alpha;j}^{+}(x), \qquad &&
(\Lambda _+ -\lambda_j^- ) \dot{\bPsi}_{\alpha;j}^{-}(x)=
\bPsi_{\alpha;j}^{-}(x), & \nonumber\\
& (\Lambda _- -\lambda_j^+ )
\dot{\bPhi}_{ir;j}^{+}(x)=\bPhi_{\alpha;j}^{+}(x), \qquad &&
(\Lambda _- -\lambda_j^- ) \dot{\bPhi}_{\alpha;j}^{-}(x)=
\bPhi_{\alpha;j}^{-}(x), & 
\end{alignat*}
i.e., $\dot{\bPsi}_{\alpha;j}^{+}(x) $ and $\dot{\bPhi}_{\alpha;j}^{+}(x) $ are adjoint eigenfunctions of
$\Lambda _+ $ and $\Lambda _- $. This means that $\lambda _j^\pm$, $j=1,\dots, N $ are also the discrete
eigenvalues of $\Lambda_\pm $ but the corresponding eigenspaces of $\Lambda _\pm $ have
double the dimensions of the ones of $L $; now they are spanned by both $\bPsi_{\mp\alpha;j}^{\pm}(x) $ and
$\dot{\bPsi}_{\mp\alpha;j}^{\pm}(x) $. Thus the sets $\{\Psi \} $ and $\{\Phi \} $ are the complete sets of
eigen- and adjoint functions of $\Lambda _+ $ and $\Lambda _- $.

Therefore the completeness relation (\ref{eq:5.23}) can be viewed as the spectral decompositions of the
recursion operators $\Lambda_\pm $. It is also obvious that the continuous spectrum of these operators
f\/ills up the real axis $\bbbr$ and has multiplicity $2n$; the discrete spectrum consists of the eigenva\-lues~$\lambda_k^\pm$ and each of them has multiplicity $2$.

\section[Resolvent and spectral decompositions in the
spinor representation of $\mathfrak{g}\simeq B_r$]{Resolvent and spectral decompositions\\ in the
spinor representation of $\boldsymbol{\mathfrak{g}\simeq B_r}$}\label{section5}

Using the general theory one can calculate the explicit form of the Cartan--Weyl
basis in the spinor representations. In Appendices \ref{appendixB} and~\ref{appendixC} below we give the results
for $r=2$ and $r=3$. Therefore in the spinor representation the Lax operators take the
form:
\begin{gather}\label{eq:L-spin}
L_{\rm sp} \psi_{\rm sp} = i \frac{\partial \psi_{\rm sp} }{\partial x} + (Q_{\rm sp} - \lambda J_{\rm sp})\psi_{\rm sp}(x,\lambda) =0,
\end{gather}
where $Q_{\rm sp}(x,t)$ and $J_{\rm sp}$ are $2^r\times 2^r$ matrices of the form:
\begin{gather*}
Q_{\rm sp}  = \left(\begin{array}{cc}  0 & \q \\ \q^\dag  & 0 \end{array}\right) , \qquad
J_{\rm sp}  = \frac{1}{2}\left(\begin{array}{cc} \openone_2 & 0 \\ 0 & -\openone_2 \end{array}\right),
\end{gather*}
where the explicit form of $\q (x)$ for $r=2$ and $r=3$ is given by equations (\ref{eq:L-sp2''}) and (\ref{eq:L-sp2}) below.

It is well known \cite{Bourb1} that the spinor representations of $so(2r+1)$ are realized by symplectic (resp. orthogonal)
matrices if $r(r+1)/2 $ is odd (resp. even). Thus in what follows we will view the spinor
representations of $so(2r+1)$ as typical representations of $sp(2^r)$ (resp. $so(2^r)$) algebra. Combined with the
corresponding  value of $J_{\rm sp}$ (\ref{eq:L-sp2''}) one can conclude that
that the Lax operator $L_{\rm sp}$  for odd values of $r(r+1)/2 $ can be related  to the C.III-type symmetric spaces.
The potential $Q_{\rm sp}(x)$ however is not a generic one; it may be obtained from the generic potential as a special reduction,
which  picks up  $so(2r+1)$  as the subalgebra of $sp(2^r)$. Below we will construct an automorphism whose kernel will
pick up $so(2r+1)$ as a subalgebra of $sp(2^r)$.

Similarly the Lax operator $L_{\rm sp}$ above for even values of $r(r+1)/2 $ can be related  to the $so(2^r)$ algebra.
The element $J_{\rm sp}$ (\ref{eq:L-sp2''})  is characteristic for the {\bf D.III}-type symmetric spaces. The potential
$Q_{\rm sp}(x)$  may be obtained from the generic potential as a special reduction,
which  picks up  $so(2r+1)$  as the subalgebra of $so(2^r)$.

The spectral problem (\ref{eq:L-spin}) is technically more simple to treat. It has the form of block-matrix AKNS
which means that the corresponding Jost solutions and FAS are determined as follows:
\begin{gather*}
\psi(x,\lambda)  \mathop{\simeq}\limits_{x\to\infty} e^{-i\lambda Jx},  \qquad \phi(x,\lambda)
\mathop{\simeq}\limits_{x\to -\infty} e^{-i\lambda Jx}, \qquad
T(\lambda)  = \left(\begin{array}{cc} \a^+ & -\b^- \\ \b^+ & \a^- \end{array}\right) ,   \\
\psi(x,\lambda)  = (\psi^-(x,\lambda), \psi^+(x,\lambda) ),  \qquad \phi(x,\lambda)   = (\phi^+(x,\lambda), \phi^-(x,\lambda) ), \\
\chi^+(x,\lambda)  = (\phi^+(x,\lambda), \psi^+(x,\lambda) ),  \qquad \chi^-(x,\lambda)   = (\psi^-(x,\lambda), \phi^-(x,\lambda) ).
\end{gather*}

\subsection{The Gauss factors in the spinor representation}\label{section5.1}

The spectral theory of the Lax operators related to the symmetric spaces of C.III and D.III types were
developed in \cite{VSG*94,82}. What is dif\/ferent here is the special choice of the rank and the
additional reduction $\pi_{\rm B_r}$ which picks up the spinor representation of $so(2r+1)$. In  our
considerations below we will assume that this reduction is applied. So though all our $2^r\times 2^r$
matrices are split into blocks of dimension $2^{r-1}\times 2^{r-1}$, the corresponding group (resp. algebraic)
elements belong to the (spinor representation of) group $SO(2r+1)$ (resp. algebra $so(2r+1)$).
Thus we def\/ine the FAS of $L_{\rm sp}$ by:
\begin{gather}
\chi_{\rm sp} ^+(x,\lambda )  \equiv  \left(|\phi ^+\rangle , |\psi
^+\hat{c}^+\rangle\right)(x,\lambda) = \phi (x,\lambda )
\S_{\rm sp}^+(\lambda ) = \psi_{\rm sp} (x,\lambda ) \T^-_{\rm sp}(\lambda )D^+_{\rm sp}(\lambda ) ,
\nonumber\\
\chi_{\rm sp} ^-(x,\lambda )  \equiv  \left(|\psi ^-\hat{c}^-\rangle ,
|\phi ^-\rangle \right)(x,\lambda ) = \phi (x,\lambda )
\S_{\rm sp}^-(\lambda )=\psi_{\rm sp} (x,\lambda )\T_{\rm sp}^+(\lambda ) D_{\rm sp}^-(\lambda ) ,\label{eq:6.3}
\end{gather}
where the block-triangular functions $\S_{\rm sp}^\pm(\lambda ) $ and
$\T_{\rm sp}^\pm(\lambda ) $ are given by:
\begin{alignat}{3}
& \S_{\rm sp}^+(\lambda ) = \left( \begin{array}{cc} \openone  &
\d^-\hat{\c}^+(\lambda )\\ 0 & \openone \end{array}\right),\qquad &&
\T_{\rm sp}^-(\lambda ) =  \left( \begin{array}{cc} \openone & 0 \\
\b^+\hat{\a}^+(\lambda ) & \openone  \end{array}\right), & \nonumber\\
& \S_{\rm sp}^-(\lambda ) =  \left( \begin{array}{cc} \openone & 0 \\
-\d^+\hat{\c}^-(\lambda ) & \openone  \end{array}\right), \qquad &&
\T_{\rm sp}^+(\lambda ) = \left( \begin{array}{cc} \openone  &
-\b^-\hat{\a}^-(\lambda ) \\ 0 & \openone \end{array}\right). & \label{eq:6.4}
\end{alignat}
The matrices $D_{\rm sp}^\pm(\lambda )$ are  block-diagonal and equal:
\begin{gather*}
D_{\rm sp}^+(\lambda ) = \left( \begin{array}{cc} \a^+(\lambda ) &0 \\ 0 &
\hat{\c}^+(\lambda ) \end{array}\right), \qquad D_{\rm sp}^-(\lambda ) =
\left(\begin{array}{cc} \hat{\c}^-(\lambda ) &0 \\ 0 &
\a^-(\lambda )
\end{array}\right).
\end{gather*}
The supper scripts $\pm$ here refer to their analyticity properties for $\lambda \in\bbbc_\pm$.

All factors $\S_{\rm sp}^\pm $, $\T_{\rm sp}^\pm $ and $D_{\rm sp}^\pm $ take values in the spinor representation of
the group $SO(2r+1)$ and are determined by the minimal sets of scattering data (\ref{eq:T_i-dress}).
Besides, since
\begin{gather*}
T_{\rm sp}(\lambda )  = \T_{\rm sp}^-(\lambda )D_{\rm sp}^+(\lambda )\hat{\S}_{\rm sp}^+(\lambda ) =
\T_{\rm sp}^+(\lambda )D_{\rm sp}^-(\lambda )\hat{\S}_{\rm sp}^-(\lambda ), \nonumber\\
\hat{T}_{\rm sp}(\lambda )  = \S_{\rm sp}^+(\lambda )\hat{D}_{\rm sp}^+(\lambda )
\hat{\T}_{\rm sp}^-(\lambda ) = \S_{\rm sp}^-(\lambda )\hat{D}_{\rm sp}^-(\lambda )
\hat{\T}_{\rm sp}^+(\lambda ), 
\end{gather*}
we can view the factors $\S_{\rm sp}^\pm $, $\T_{\rm sp}^\pm $ and $D_{\rm sp}^\pm $ as
generalized Gauss decompositions (see \cite{Helg}) of~$T_{\rm sp}(\lambda )$ and its inverse.

From equations (\ref{eq:6.3}), (\ref{eq:6.4}) one can derive:
\begin{alignat}{3}
& \chi_{\rm sp} ^+(x,\lambda )  =  \chi_{\rm sp} ^-(x,\lambda ) G_{0,\rm sp}(\lambda ), \qquad &&
 \chi_{\rm sp} ^-(x,\lambda ) = \chi_{\rm sp} ^+(x,\lambda )\hat{G}_{0,\rm sp}(\lambda ), & \label{eq:9.4} \\
& G_{0,\rm sp}(\lambda ) = \left(\begin{array}{cc} \openone
& \tau^+ \\ \tau^- & \openone + \tau^-\tau^+ \\ \end{array}
\right), \qquad && \hat{G}_{0,\rm sp}(\lambda ) = \left(\begin{array}{cc}
\openone + \tau^+\tau^- & -\tau^+ \\ -\tau^- & \openone
\\ \end{array} \right)& \label{eq:11.3}
\end{alignat}
valid for $\lambda \in \bbbr$. Below  we introduce:
\begin{gather}\label{eq:12.3}
X_{\rm sp}^\pm(x,\lambda ) = \chi_{\rm sp} ^\pm(x,\lambda ) \re^{i\lambda Jx}.
\end{gather}
Strictly speaking it is $X_{\rm sp}^\pm(x,\lambda)$ that allow analytic
extension for $\lambda \in \bbbc_\pm$. They have also another nice
property, namely their asymptotic behavior for $\lambda \to\pm\infty  $ is given by:
\begin{gather}\label{eq:12.1}
\lim_{\lambda \to\infty } X_{\rm sp}^\pm(x,\lambda ) =\openone .
\end{gather}
Along with $X_{\rm sp}^\pm(x,\lambda)$ we can use another set of FAS
$\tilde{X}_{\rm sp}^\pm(x,\lambda)= X_{\rm sp}^\pm(x,\lambda)\hat{D}_{\rm sp}^\pm$, which
also satisfy equation (\ref{eq:12.1}) due to the fact that:
\begin{gather*}
\lim_{\lambda \to\infty } D_{\rm sp}^\pm(\lambda ) =\openone.
\end{gather*}
The equations (\ref{eq:9.4}) and (\ref{eq:11.3}) can be written down as:
\begin{gather}\label{eq:12.4}
X_{\rm sp}^+(x,\lambda ) = X_{\rm sp}^-(x,\lambda ) G_{\rm sp}(x,\lambda ), \qquad \lambda
\in \bbbr
\end{gather}
with
\begin{gather*}
\tilde{G}_{\rm sp}(x,\lambda ) = \re^{-i\lambda Jx} \tilde{G}_{0,\rm sp}(\lambda
)e^{i\lambda Jx}, \qquad  \tilde{G}_{0,\rm sp}(\lambda ) =
\left(\begin{array}{cc} \openone +\rho^-\rho^+ & \rho^- \\
\rho^+ & \openone \\ \end{array} \right).
\end{gather*}
Equations (\ref{eq:12.4})  combined with (\ref{eq:12.1}) are known as a
Riemann--Hilbert problem (RHP) with~ca\-nonical normalization \cite{Gakhov}. It has
unique regular solution; the matrix-valued solutions $X_{0,\rm sp}^+(x,\lambda ) $
and $X_{0,\rm sp}^-(x,\lambda ) $ of (\ref{eq:12.4}), (\ref{eq:12.1}) is
called regular if $\det X_{0,\rm sp}^\pm(x,\lambda ) $ does not vanish for
any $\lambda \in\bbbc_\pm $.

One can derive the following integral decomposition for $X_{\rm sp}^\pm(x,\lambda ) $:
\begin{gather}\label{eq:X+}
X_{\rm sp}^+(x,\lambda ) = \openone + {1 \over 2\pi \ri} \int_{-\infty
}^{\infty } {d\mu\over \mu -\lambda } X_{\rm sp}^-(x,\mu ) K_1(x,\mu ) +
\sum_{j =1}^{N} \frac{X_{j,\rm sp}^-(x)K_{1,j}(x)}{\lambda _j^- -\lambda},
\\
\label{eq:X-}
X_{\rm sp}^-(x,\lambda ) = \openone + {1 \over 2\pi \ri} \int_{-\infty
}^{\infty } {d\mu\over \mu -\lambda } X_{\rm sp}^-(x,\mu ) K_2(x,\mu ) -
\sum_{j =1}^{N} \frac{ X_{j,\rm sp}^+(x)K_{2,j}(x)}{\lambda _j^+ -\lambda},
\end{gather}
where $X_{j,\rm sp}^\pm(x)=X_{\rm sp}^\pm(x,\lambda_j^\pm)$ and
\begin{gather*}
K_{1,j}(x) = \re^{-\ri \lambda_j^- J x} \left(\begin{array}{cc} 0
& \rho_j^+ \\ \tau_j^- & 0 \\ \end{array} \right)  \re^{\ri
\lambda_j^- J x} , \qquad K_{2,j}(x) = \re^{-\ri \lambda_j^+ J x}
\left(\begin{array}{cc} 0 & \tau_j^+ \\ \rho_j^- & 0 \\
\end{array} \right)\re^{\ri \lambda_j^+ J x}.
\end{gather*}
Equations (\ref{eq:X+}), (\ref{eq:X-}) can be viewed as a set of
singular integral equations which are equivalent to the RHP. For
the MNLS these were f\/irst derived in~\cite{Ma1}.

Finally,   the potential $Q(x,t)$ can be recovered from the solutions
$X_{\rm sp}^\pm(x,\lambda)$ of RHP (\ref{eq:12.4})   with a canonical normalisation
(\ref{eq:12.1}). Skipping the details, we provide here only the f\/inal result:
\begin{gather*}
Q_{\rm sp}(x,t) = \lim_{\lambda\to\infty} \lambda (J - X_{\rm sp}^\pm(x,\lambda) J
\hat{X}_{\rm sp}^\pm(x,\lambda)]) =[J, X_1(x)],
\end{gather*}
where $X_1(x) = \lim\limits_{\lambda\to\infty} (X_{\rm sp} (x,\lambda)-\openone)$.

\subsection[Definition and properties of $R_{\rm sp}^\pm(x,y,\lambda)$]{Def\/inition and properties of $\boldsymbol{R_{\rm sp}^\pm(x,y,\lambda)}$}\label{section5.2}

The resolvent $R_{\rm sp}(\lambda ) $ of $L_{\rm sp}$ is again expressed through the FAS  in the form:
\begin{gather*}
R_{\rm sp}(\lambda ) f(x) = \int_{-\infty }^{\infty } R_{\rm sp}(x,y,\lambda ) f(y),
\end{gather*}
where $R_{\rm sp}(x,y,\lambda ) $ are given by:
\begin{gather*}
R_{\rm sp}(x,y,\lambda ) = \left\{ \begin{array}{ll}
R_{\rm sp}^+(x,y,\lambda ) \ & \mbox{for} \ \lambda \in \bbbc^+, \\
R_{\rm sp}^-(x,y,\lambda ) \ & \mbox{for} \ \lambda \in \bbbc^-,
\end{array} \right.
\end{gather*}
and
\begin{gather*}
R_{\rm sp}^\pm (x,y,\lambda )=\pm i\chi_{\rm sp} ^\pm(x,\lambda ) \Theta ^\pm (x-y)
\hat{\chi }_{\rm sp}^\pm (y,\lambda ), \qquad
\Theta ^\pm(z) = \left(\begin{array}{cc} \theta (\mp z) \openone & 0 \\ 0 & -\theta (\pm z) \openone
\end{array}\right).
\end{gather*}

\subsection{The dressing factors in the spinor representation}\label{section5.3}

The asymptotics of the dressing factor for $x\to\pm\infty$ are:
\begin{gather*}
u_\pm = \exp \left( \ln c_1(\lambda) J_{\rm sp}\right) ,
\end{gather*}
where
\begin{gather*}
u_+ = \left(\begin{array}{cc} \sqrt{c_1} \openone_{2^r} & 0 \\
0 & 1/\sqrt{c_1} \openone_{2^{r-1}} \end{array}\right) , \qquad u_- = \left(\begin{array}{cc} 1/\sqrt{c_1} \openone_{2^{r-1}}& 0 \\
0 & \sqrt{c_1} \openone_{2^{r-1}} \end{array}\right) .
\end{gather*}
Thus the asymptotics of the projectors $P$ and $\bar{P}$ in the spinor representation
are projectors of rank $2^{r-1}$. Since the dynamics of the MNLS does not change the rank of the projectors we
conclude that  the dressing factor in the spinor representation must be of the form:
\begin{gather*}
u(x,\lambda) = \exp \left( \frac{1}{2}\ln c_1(\lambda) (P_1 - \bar{P}_1)\right) =
\sqrt{c_1(\lambda)} P_1(x,t) + \frac{1}{\sqrt{c_1(\lambda)}} \bar{P}_1(x,t)
\end{gather*}
dressing factors  of such form with non-rational dependence on $\lambda$ were considered for the f\/irst time in
by Ivanov \cite{I04} for the MNLS related to symplectic algebras. Here we see that such construction can be
used also for the orthogonal algebras.

However when it comes to analyze the relations between the `naked' and the dressed scattering matrices and their
Gauss factors we get integer powers of $c_1(\lambda)$. Indeed, for the simplest case when the dressing procedure
is applied just once we have:
\begin{alignat*}{3}
& T_{\rm sp}(\lambda) = \hat{u}_+ T_{0,\rm sp}(\lambda) u_-(\lambda), \qquad && D_{\rm sp}^\pm(\lambda) =
\hat{u}_+ D_{0,\rm sp}^\pm (\lambda) u_-(\lambda),& \nonumber\\
& T_{\rm sp}^\pm(\lambda)  = \hat{u}_+ T^\pm _{0,\rm sp}(\lambda) u_+(\lambda), \qquad &&
S_{\rm sp}^\pm(\lambda) = \hat{u}_- S_{0,\rm sp}^\pm (\lambda) u_-(\lambda),&
\end{alignat*}
and as a consequence
\begin{alignat*}{3}
& \a^+_{\rm sp}(\lambda) = \frac{1}{c_1^+(\lambda)} \a^+_{0,\rm sp}(\lambda), \qquad&&
\a^-_{\rm sp}(\lambda) = c_1^+(\lambda) \a^-_{0,\rm sp}(\lambda), &\nonumber \\
& \rho^+_{\rm sp}(\lambda) = \frac{1}{c_1^+(\lambda)} \rho^+_{0,\rm sp}(\lambda), \qquad&&
\rho^-_{\rm sp}(\lambda) = c_1^+(\lambda) \rho^-_{0,\rm sp}(\lambda),&\nonumber \\
& \tau^+_{\rm sp}(\lambda) = \frac{1}{c_1^+(\lambda)} \tau^+_{0,\rm sp}(\lambda), \qquad&&
\tau^-_{\rm sp}(\lambda)= c_1^+(\lambda) \tau^-_{0,\rm sp}(\lambda). &
\end{alignat*}

\subsection[The spectral decompositions of $L_{\rm sp}$]{The spectral decompositions of $\boldsymbol{L_{\rm sp}}$}\label{section5.4}

Again apply the contour integration method to the kernel $R_{\rm sp}^\pm $ and derive the following
completeness relation:
\begin{gather}
\delta(x-y)  \openone_{2^r}
  = {1\over  2\pi}\int_{-\infty}^\infty d\lambda\left\{  |\phi^+(x,\lambda)\rangle \hat{\a}^+(\lambda) \langle \psi^+(y,\lambda)|
 -|\phi^-(x,\lambda)\rangle \hat{\a}^-(\lambda) \langle \psi^-(y,\lambda)|  \right\} \nonumber\\
 \phantom{\delta(x-y)  \openone_{2^r}=}{} +  \sum_{j=1}^N
\left( \Res_{\lambda=\lambda_j^+}R^+(x,y) +  \Res_{\lambda=\lambda_j^-}R^-(x,y) \right).\label{eq:3.44'}
\end{gather}
The residues in (\ref{eq:3.44'}) can be expressed by the properly normalized eigenfunctions of
$L_{\rm sp}$ corresponding to the eigenvalues  $\lambda_j^\pm$.

Thus we conclude that the continuous spectrum of $L_{\rm sp}$ f\/ills up $\bbbr$ and has multiplicity $2^r$.
The discrete eigenvalues $\lambda_k^\pm$ are simple poles of the resolvent.

\section[Dressing factors and higher representations of $\mathfrak{g}$]{Dressing factors and higher representations of $\boldsymbol{\mathfrak{g}}$}\label{section6}

Here we will brief\/ly outline how, starting from equation (\ref{eq:rank1}) one can construct the
dressing factors in any irreducible representation of the Lie algebra $\mathfrak{g} $. We will
illustrate this on one of the simple nontrivial examples of $u(x,\lambda)$ with rank-1 projectors $P(x)$
and $\bar{P}(x)$. Our intention is to outline the explicit $\lambda$-dependence of $u(x,\lambda)$
in any IRREP. To this end it will be most convenient to use the f\/irst line of equation (\ref{eq:upm1}):
\begin{gather}\label{eq:upm2}
u(x,\lambda)= \exp \left( \ln c_1(\lambda)(P_1 - \bar{P}_1)\right).
\end{gather}
Note that $P(x)-\bar{P}(x) \in \mathfrak{g} $ and therefore the right hand side of (\ref{eq:upm2})
will be an a Lie group element.
In what follows we will assume that $\rank P(x)=\rank \bar{P}(x)=1$ and will derive explicitly the
$\lambda$-dependence of the right hand side of equation (\ref{eq:upm2}) in any irreducible representation of
$\mathfrak{g} $.

In fact it will be enough to analyze the $\lambda$-dependence of the asymptotic of $u(x,\lambda)$ for $x\to\pm\infty$.
\begin{gather*}
\lim_{x\to\infty} (P(x)-\bar{P}(x)) = H_{e_1}.
\end{gather*}
In order to be more specif\/ic we will do our considerations for the case when $\mathfrak{f}\simeq so(7) $. This algebra
is of rank~3. We also choose the representation with highest weight
\begin{gather*}
\omega = 3\omega_3 = \frac{3}{2} (e_1+e_2+e_3).
\end{gather*}
The structure of the weight system $\Gamma ^{(\omega)} $ is described in Table~\ref{tab:1}.

\begin{table}[t]\centering

\caption{The structure of the weight system $\Gamma ^{(3\omega_1)} $ for the algebra $so(7)$ with
dimension 112. We list here the number of weights of dif\/ferent lengths $\ell (\gamma)$ and their multiplicity $\mu(\gamma)$
and length. The indices $i$, $j$, $k$ are dif\/ferent and take values $1$, $2$ and $3$.\label{tab:1}}

\vspace{1mm}

\begin{tabular}{|c||c|c|c|c|}
  \hline
 weight type & $\Gamma_1$ & $\Gamma_2$ & $\Gamma_3$ & $\Gamma_4$ \\ \hline
  & $\frac{3}{2} (\pm e_1 \pm e_2 \pm e_3)$ & $\frac{1}{2} (\pm 3e_i \pm 3e_j \pm e_k)$ &
$ \frac{1}{2} (\pm 3e_i \pm e_j \pm e_k)$ & $\frac{1}{2} (\pm e_1 \pm e_2 \pm e_3)$\tsep{2pt}\bsep{2pt} \\ \hline
\# $(\gamma) $ & 8 & 24 & 24 & 24 \\  $\mu(\gamma) $ & 1 & 1 & 2 & 4 \\
$\ell (\gamma)$ & $\frac{27}{4}$ & $\frac{19}{4} $  & $\frac{11}{4} $  & $\frac{3}{4} $ \bsep{2pt} \\ \hline
\end{tabular}

\end{table}

It is natural to expect that $u(x,\lambda)$ will have the same type of $\lambda$-dependence \cite{PLA126} as its asymptotic
for $x\to\pm\infty$. Therefore we will evaluate the right hand side of equation (\ref{eq:upm2}) for $x\to\infty$. Doing this we
well use the known formula
\begin{gather*}
H_{e_1} = \sum_{\gamma\in\Gamma ^{(3\omega_3)} } (e_1,\gamma) |\gamma\rangle \langle \gamma|.
\end{gather*}
Therefore we have to arrange the weights in $\Gamma ^{(3\omega_3)}$ according to their scalar products with $e_1$,
namely:
\begin{gather*}
\Gamma ^{(3\omega_3)} = \Gamma_{3/2} \cup \Gamma_{1/2}\cup \Gamma_{-1/2} \cup \Gamma_{-3/2}  , \nonumber\\
\Gamma_{3/2}  \equiv \left\{ \frac{3}{2} ( e_1 \pm e_2 \pm e_3) \right\} \cup \left\{\frac{1}{2} (3 e_1 \pm 3e_j \pm e_k)   \right\}
\cup \left\{ \frac{1}{2} ( 3e_1 \pm e_2 \pm e_3) \right\} ,\nonumber\\
\Gamma_{1/2}  \equiv \left\{\frac{1}{2} ( e_1 \pm 3e_2 \pm 3e_3) \right\}  \cup \left\{ \frac{1}{2} ( e_1 \pm 3e_j \pm e_k) \right\}\cup
\left\{ \frac{1}{2} ( e_1 \pm e_2 \pm e_3) \right\} , \nonumber\\
\Gamma_{-1/2}  \equiv \left\{\frac{1}{2} ( -e_1 \pm 3e_2 \pm 3e_3) \right\}  \cup \left\{ \frac{1}{2} ( -e_1 \pm 3e_j \pm e_k) \right\}\cup
\left\{ \frac{1}{2} ( -e_1 \pm e_2 \pm e_3) \right\} , \nonumber\\
\Gamma_{-3/2}  \equiv \left\{ \frac{3}{2} ( -e_1 \pm e_2 \pm e_3) \right\} \cup \left\{\frac{1}{2} (-3 e_1 \pm 3e_j \pm e_k)   \right\}
\cup \left\{ \frac{1}{2} ( -3e_1 \pm e_2 \pm e_3) \right\} ,
\end{gather*}
where $j$, $k$ take the values $2$ and $3$. As a result we obtain
\begin{gather}\label{eq:He2}
H_{e_1} = \frac{3}{2} \pi_{3/2} +  \frac{1}{2} \pi_{1/2} -  \frac{1}{2} \pi_{-1/2} - \frac{3}{2} \pi_{-3/2} ,
\end{gather}
where the projectors $\pi_a$, $a=\pm \frac{3}{2}, \pm \frac{1}{2}$ are equal to
\begin{gather*}
\pi_a = \sum_{\gamma \in \Gamma_a}^{} |\gamma\rangle \langle \gamma|,
\end{gather*}
and obviously satisfy the relations:
\begin{gather*}
\pi_a \pi_b = \delta_{ab} \pi_a, \qquad \rank \pi_{3/2} = \rank \pi_{-3/2} = 20,  \qquad \rank \pi_{1/2} = \rank \pi_{-1/2} = 36,
\end{gather*}
and $ \pi_{3/2} + \pi_{1/2} + \pi_{-1/2} +\pi_{-3/2} = \openone_{112}  $.
Thus the $Q(x)-\lambda J$ acquires the following block-matrix form:
\begin{gather*}
\left(\begin{array}{cccc}
\frac{3}{2}\lambda \openone_{20} & Q_{(12)} & Q_{(13)} & Q_{(14)} \vspace{1mm}\\ Q_{(21)} & \frac{1}{2} \lambda\openone_{36} &
Q_{(23)} & Q_{(24)} \vspace{1mm}\\ Q_{(31)} & Q_{(32)} & -\frac{1}{2} \lambda\openone_{36} & Q_{(34)} \vspace{1mm}\\ Q_{(41)} & Q_{(42)} & Q_{(43)} &
-\frac{3}{2} \lambda\openone_{20} \end{array}\right).
\end{gather*}
Formally this potential can be viewed as related to  the homogeneous space $SO(112)/S(O(40)\otimes O(72))$. However
all matrix elements of the potential $Q(x)$ are determined by the f\/ive components of the vector $\vec{q}$ and their
complex conjugate. This deep reduction imposed on $Q(x)$ corresponds to the fact that instead of considering a generic
element of this homogeneous space, we rather pick up the representation of $so(7)$ with highest weight $3\omega_3$.

Inserting equation (\ref{eq:He2}) into equation (\ref{eq:upm2}) we get:
\begin{gather*}
\lim _{x\to\infty} u^{(3\omega_3)} (x,\lambda) = (c(\lambda))^{3/2} \pi_{3/2} + (c(\lambda))^{1/2} \pi_{1/2}
+  (c(\lambda))^{-1/2} \pi_{-1/2} +  (c(\lambda))^{-3/2} \pi_{-3/2},
\end{gather*}
and as a result for the $\lambda$-dependence of $ u^{(3\omega_3)} (x,\lambda)$ and its inverse we get:
\begin{gather*}
u^{(3\omega_3)} (x,\lambda) = (c(\lambda))^{3/2} \pi_{3/2}(x) + (c(\lambda))^{1/2} \pi_{1/2}(x)\\
\phantom{u^{(3\omega_3)} (x,\lambda) =}{}
+  (c(\lambda))^{-1/2} \pi_{-1/2}(x) +  (c(\lambda))^{-3/2} \pi_{-3/2} (x), \\
 (u^{(3\omega_3)})^{-1} (x,\lambda) = (c(\lambda))^{-3/2} \pi_{3/2}(x) + (c(\lambda))^{-1/2} \pi_{1/2}(x)\\
\phantom{(u^{(3\omega_3)})^{-1} (x,\lambda) =}{}
+  (c(\lambda))^{1/2} \pi_{-1/2}(x) +  (c(\lambda))^{3/2} \pi_{-3/2} (x).
\end{gather*}
The four projectors $\pi_a(x)$ have the same properties as their asymptotic values:
\begin{gather*}
\pi_a(x) \pi_b (x) = \delta_{ab} \pi_a (x), \\
\rank \pi_{3/2}(x) = \rank \pi_{-3/2}(x) = 20,   \qquad \rank \pi_{1/2}(x)  = \rank \pi_{-1/2}(x) = 36.
\end{gather*}
Their explicit $x$ dependence as well as the interrelation between the potentials $Q_{(0)}(x)$  and $Q_{(1)}(x)$ follow
from the equation for $u(x,\lambda)$ (\ref{eq:u-eq}) considered in the representation $V^{(3\omega_3)} $.
In particular we get:
\begin{gather*}
Q_{(1)}(x) - Q_{(0)}(x) = \lim_{\lambda\to\infty} \lambda \left( J - u^{(3\omega_3)}  J (u^{(3\omega_3)})^{-1} (x,\lambda) \right) \\
\phantom{Q_{(1)}(x) - Q_{(0)}(x)}{} = (\lambda^+ -\lambda^-) \left[ J, \frac{3}{2} (\pi_{3/2}(x) -\pi_{-3/2}(x)) + \frac{1}{2} (\pi_{1/2}(x)
 -\pi_{-1/2}(x))\right].
\end{gather*}
Though the expressions for the dressing factor in this representation seem to be rather complex,
nevertheless they are determined uniquely by the projector $P_1(x)$, or equivalently, through the
polarization vector $|n_1(x)\rangle $. The corresponding expressions can be using the fact that
$V^{(3\omega_3)}$ can be extracted as the invariant subspace of the tensor product $V^{(\omega_3)}\otimes
V^{(\omega_3)}\otimes V^{(\omega_3)}$ corresponding to the highest weight vector $3\omega_3$.
Therefore one can conclude that the matrix elements of the projectors $\pi_a$, $a=\pm 3/2, \pm 1/2$
will be polynomials of sixth order of the components of $|n_1(x)\rangle$.

Our f\/inal remark concerns the analyticity properties of the FAS
dressed by $u^{(3\omega_3)} (x,\lambda)$. The factor itself
contains powers of root square of $c(\lambda)$ which in general
may lead to essential singularities at $\lambda =\lambda^+$ and
$\lambda =\lambda^-$. Note however that the dressed FAS in this
representation are obtained from the regular solutions via the
analog of equation (\ref{eq:Dressfactor}):
\begin{gather}
\chi^{\pm, 3\omega_3}_1(x,t,\lambda)=u^{(3\omega_3)}(x,\lambda) \chi^{\pm,3\omega_3}_0(x,t,\lambda)
(u_{-}^{3\omega_3})^{-1}(\lambda ), \nonumber\\
 u_-^{3\omega_3}(\lambda )  =\lim_{x\to -\infty }u^{3\omega_3}(x,\lambda ) =
 \exp \left( -\ln c_1(\lambda)H_{e_1}\right) \nonumber\\
\phantom{u_-^{3\omega_3}(\lambda )}{} =(c(\lambda))^{3/2} \pi_{=3/2} + (c(\lambda))^{1/2} \pi_{-1/2}
+  (c(\lambda))^{-1/2} \pi_{1/2} +  (c(\lambda))^{-3/2} \pi_{3/2}.\label{eq:Dressfac3}
\end{gather}
It is not dif\/f\/icult to check that the right hand side of f\/irst line in equation (\ref{eq:Dressfac3}) does
not contain square root terms of $c(\lambda)$; all half-integer powers of $c(\lambda)$ get multiplied
by other half-integer powers and the result is that $\chi^{\pm, 3\omega_3}_1(x,t,\lambda)$
acquires additional pole singularities at $\lambda =\lambda^+$ and $\lambda =\lambda^-$.

\section{Conclusions}\label{section7}

We have analyzed the spectral properties of the Lax operators related to three
dif\/ferent representations of $\mathfrak{g}\simeq B_r $: the typical, the adjoint and
the spinor representation. In all these cases the spectral properties such as: i) the
multiplicity of the continuous spectra and of the discrete eigenvalues; ii) the explicit
form of the dressing factors; iii) the completeness relations of the eigenfunctions are
substantially dif\/ferent. However the minimal sets of scattering data $\mathfrak{T}_i $
are provided by the same sets of functions, i.e.\ the sets $\mathfrak{T}_i $ are invariant with
respect to the choice of the representation of $\mathfrak{g} $.

Our considerations were performed for the class of smooth potentials $Q(x)$ vanishing fast enough
for $x\to\pm\infty$. Similar results can be derived also for the class of potentials tending to
constants $Q_\pm$ for $x\to\pm\infty$.

\appendix

\section[The adjoint representations of $so(5)$ and  $so(7)$]{The adjoint representations of $\boldsymbol{so(5)}$ and  $\boldsymbol{so(7)}$}\label{appendixA}

The adjoint representations of all orthogonal algebras $so(2r+1)$ and $so(2r)$ are characterized by the
fundamental weight $\omega_2=e_1+e_2$. By def\/inition the corresponding weight system is
\begin{gather*}
\Gamma_{\omega_2} \equiv \Delta \cup \{0\}_r,
\end{gather*}
where $\Delta$ is the root system of the algebra and $\{0\}$ is a vanishing weight with multiplicity~$r$. We will order the weights in $\Gamma_{\omega_2}$ according to their scalar products with~$e_1$.

Let us consider in more detail the two special cases of $so(2r+1)$ with $r=2$ and $r=3$. For $r=2$ the
adjoint representation is 10-dimensional. Ordering the roots as mentioned above we get:
\begin{gather*}
\Delta_1^+ \simeq \{ e_1+e_2, e_1,  e_1-e_2\}, \qquad \Delta_0 \simeq \{ e_2, -e_2 \}, \qquad
\Delta_1^- \simeq \{ -e_1+e_2, e_1,  -e_1-e_2\},
\end{gather*}
As a result the element $J$ in the adjoint representation takes the form:
\begin{gather*}
J_{\rm ad} = \left(\begin{array}{ccc} \openone_2 & 0 & 0 \\ 0 & 0_6 & 0 \\ 0 & 0 & -\openone_2 \end{array}\right), \qquad
Q_{\rm ad} = \left(\begin{array}{ccc} 0 & Q_{\rm ad; 12} & 0 \\ Q_{\rm ad; 21} & 0_6 & Q_{\rm ad; 23} \\ 0 & Q_{\rm ad; 32} & 0 \end{array}\right).
\end{gather*}
Analogously for $r=3$ the adjoint representation is 21-dimensional. Ordering the roots as mentioned above we get:
\begin{gather*}
\Delta_1^+ \simeq \{ e_1+e_2, e_1+e_3, e_1, e_1-e_3,  e_1-e_2\}, \qquad \Delta_0 \simeq \{  e_2\pm e_3, -( e_2\pm e_3) \}, \\
\Delta_1^- \simeq \{ -e_1+e_2, -e_1+e_3, e_1, -e_1-e_3, -e_1-e_2\}.
\end{gather*}
As a result the element $J$ in the adjoint representation takes the form:
\begin{gather*}
J_{\rm ad} = \left(\begin{array}{ccc} \openone_5 & 0 & 0 \\ 0 & 0_{11} & 0 \\ 0 & 0 & -\openone_5 \end{array}\right), \qquad
Q_{\rm ad} = \left(\begin{array}{ccc} 0 & Q_{\rm ad; 12} & 0 \\ Q_{\rm ad; 21} & 0_{11} & Q_{\rm ad; 23} \\ 0 & Q_{\rm ad; 32} & 0 \end{array}\right).
\end{gather*}
It is not dif\/f\/icult to write down the explicit form of $Q_{\rm ad} $ but it will not be necessary.
We have used above a more compact realization of $Q_{\rm ad} $ as ${\rm ad}_Q$.

\section[The spinor representation of $so(5)$]{The spinor representation of $\boldsymbol{so(5)}$}\label{appendixB}

The highest weight and the weight system of $so(5)$ are given by \cite{Bourb1,Helg}. It is well known that
$so(5)\simeq sp(4)$ so the spinor representation of $so(5)$ is realized through symplectic $sp(4)$
matrices
\begin{gather*}
\omega_2 \equiv \gamma_1  = \frac{1}{2} (e_1+e_2),  \qquad \gamma_2  = \frac{1}{2} (e_1-e_2),\qquad
\gamma_3  = -\gamma_2,  \qquad \gamma_4 = - \gamma_1.
\end{gather*}
The Cartan--Weyl basis of $so(5)$ is given by
\begin{alignat*}{3}
& E_{e_1-e_2}  = \Gamma_{2,\bar{2}} =\mathcal{E}_{2\epsilon_2}, \qquad && E_{e_1+e_2}  = \Gamma_{1,\bar{1}}  =\mathcal{E}_{2\epsilon_1}, & \\
& E_{e_1}  = \Gamma_{1,\bar{2}} -\Gamma_{2,\bar{1}} =\mathcal{E}_{\epsilon_1 + \epsilon_2}, \qquad &&
E_{e_2} = \Gamma_{1,2} -\Gamma_{2,\bar{1}} =\mathcal{E}_{\epsilon_1-\epsilon_2}, & \\
& H_{e_1} = \frac{1}{2}( \Gamma_{1,1}+\Gamma_{2,2}),  \qquad && H_{e_2}  = \frac{1}{2}( \Gamma_{1,1}-\Gamma_{2,2} ), &
\end{alignat*}
where $\bar{k}=5-k$ and
\begin{gather*}
\Gamma_{k,p} = |\gamma_k\rangle \langle \gamma_p|, \qquad 1 \leq k \leq p \leq 4.
\end{gather*}
By $\mathcal{E}_{\epsilon_i\pm \epsilon_j} $ above we have denoted the Weyl generators of $sp(4)$.
The Lax operator $L$ in the spinor representation of $so(5)$ takes the form:
\begin{gather*}
L_{\rm sp} \psi_{\rm sp} = i \frac{\partial \psi_{\rm sp} }{\partial x} + (Q_{\rm sp} - \lambda J_{\rm sp})\psi_{\rm sp}(x,\lambda) =0,
\end{gather*}
where $Q_{\rm sp}(x,t)$ and $J_{\rm sp}$ are $4\times 4$ symplectic matrices of the form:
\begin{gather}\label{eq:L-sp2''}
Q_{\rm sp}  = \left(\begin{array}{cc}  0 & \q \\ \q^\dag  & 0 \end{array}\right) , \qquad
J_{\rm sp}  = \frac{1}{2}\left(\begin{array}{cc} \openone_2 & 0 \\ 0 & -\openone_2 \end{array}\right), \qquad
\q(x,t)  = \left(\begin{array}{cc}  q_0 & q_{\bar{1}} \\ q_1  & -q_0 \end{array}\right).
\end{gather}

\section[The spinor representation of $so(7)$]{The spinor representation of $\boldsymbol{so(7)}$}\label{appendixC}

The highest weight and the weight system of $so(7)$ are given by \cite{Bourb1,Helg}
\begin{alignat*}{3}
& \omega_3 \equiv \gamma_1 = \frac{1}{2} (e_1+e_2+e_3), \qquad && \gamma_2 = \frac{1}{2} (e_1+e_2-e_3),& \\
& \gamma_3 = \frac{1}{2} (e_1-e_2+e_3),  \qquad && \gamma_4 = \frac{1}{2} (e_1-e_2-e_3), \\
& \gamma_5 = -\gamma_4, \qquad \gamma_6 = - \gamma_3, \qquad && \gamma_7 =-\gamma_2, \qquad \gamma_8=-\gamma_1.&
\end{alignat*}
The Cartan--Weyl basis of $so(7)$ is given by
\begin{alignat*}{3}
& E_{e_1-e_2}  = \Gamma_{3,\bar{4}} =\mathcal{E}_{\epsilon_3+\epsilon_4}, \qquad  && E_{e_2-e_3}  = \Gamma_{2,3} =\mathcal{E}_{\epsilon_2-\epsilon_3},& \\
& E_{e_1-e_3}  = \Gamma_{2,\bar{4}} =\mathcal{E}_{\epsilon_2+\epsilon_4}, \qquad && E_{e_1+e_2}  = \Gamma_{1,\bar{2}}  =\mathcal{E}_{\epsilon_2+\epsilon_4}, & \\
& E_{e_1+e_3}  = \Gamma_{1,\bar{3}} =\mathcal{E}_{\epsilon_1+\epsilon_3}, \qquad && E_{e_2+e_3} = \Gamma_{1,4}=\mathcal{E}_{\epsilon_1-\epsilon_4},&\\
&E_{e_1}  = \Gamma_{1,\bar{4}} +\Gamma_{2,\bar{3}} =\mathcal{E}_{\epsilon_1 + \epsilon_4} + \mathcal{E}_{\epsilon_2+\epsilon_3},  \qquad &&
 E_{e_2} = \Gamma_{1,3} +\Gamma_{2,4} =\mathcal{E}_{\epsilon_1-\epsilon_4}+\mathcal{E}_{\epsilon_2-\epsilon_4}, & \\
&E_{e_3}  = \Gamma_{1,2} -\Gamma_{3,4} =\mathcal{E}_{\epsilon_1-\epsilon_2}-\mathcal{E}_{\epsilon_3-\epsilon_4},  \qquad &&
  H_{e_1} = \frac{1}{2}( \Gamma_{1,1}+\Gamma_{2,2}+\Gamma_{3,3}+\Gamma_{4,4}),& \\
&  H_{e_1} = \frac{1}{2}( \Gamma_{1,1}+\Gamma_{2,2}-\Gamma_{3,3} -\Gamma_{4,4}), \qquad && H_{e_1}  = \frac{1}{2}( \Gamma_{1,1}-\Gamma_{2,2}+\Gamma_{3,3}-\Gamma_{4,4}),&
\end{alignat*}
where $\bar{k}=9-k$ and
\begin{gather*}
\Gamma_{k,p} = |\gamma_k\rangle \langle \gamma_p| -(-1)^{k+p}  |\gamma_{\bar{p}}\rangle \langle \gamma_{\bar{k}}|, \qquad 1 \leq k \leq p \leq 4, \\
\Gamma_{k,\bar{p}} = |\gamma_k\rangle \langle \gamma_p| +(-1)^{k+p}  |\gamma_{\bar{p}}\rangle \langle \gamma_{\bar{k}}|, \qquad 1 \leq k \leq p \leq 4.
\end{gather*}
Note that the typical representation of $so(8)$ is also $8$-dimensional. So by $\mathcal{E}_{\epsilon_i\pm \epsilon_j} $ above we have denoted
the Weyl generators of $so(8)$. So we can also consider the spinor representation of $so(7)$ as an embedding of $so(7)$ into $so(8)$. Such embedding
can be realized by taking the average of the Cartan--Weyl basis of $so(8)$ with respect to the $\bbbz_2$ external automorphism of $so(8)$ (the mirror ref\/lection) which changes $\epsilon_4 \leftrightarrow -\epsilon_4$.
In short the Lax operator $L$ in the spinor representation of $so(7)$ take the form:
\begin{gather*}
L_{\rm sp} \psi_{\rm sp} = i \frac{\partial \psi_{\rm sp} }{\partial x} + (Q_{\rm sp} - \lambda J_{\rm sp})\psi_{\rm sp}(x,\lambda) =0,
\end{gather*}
where $Q_{\rm sp}(x,t)$ and $J_{\rm sp}$ are $8\times 8$ matrices of the form:
\begin{gather}\label{eq:L-sp2}
Q_{\rm sp}  = \left(\begin{array}{cc}  0 & \q \\ \q^\dag  & 0 \end{array}\right) , \qquad
J_{\rm sp}  = \frac{1}{2}\left(\begin{array}{cc} \openone_4 & 0 \\ 0 & -\openone_4 \end{array}\right), \qquad
\q(x,t)  = \left(\begin{array}{cccc}  q_0 & q_{\bar{2}} & q_{\bar{1}} & 0 \\ q_2 & q_0 & 0  & -q_{\bar{1}} \\
q_1 & 0 & -q_0 & q_{\bar{2}} \\ 0 & - q_1 & q_2 & -q_0 \end{array}\right) .
\end{gather}
Our f\/inal remark here concerns the spinor representation with highest weight $\omega =3\omega_3$. Then
\begin{gather*}
\Gamma \simeq \left\{ \frac{3}{2} (\pm e_1\pm e_2 \pm e_3 ) ,  \frac{1}{2} (\pm e_1\pm e_2 \pm e_3 ) \right\}.
\end{gather*}
The dimension of this representation is $8$. The explicit form of the element $J$ in this representation is
$J=\diag(\frac{3}{2}\openone_{}, \frac{1}{2}\openone_{}, -\frac{3}{2}\openone_{}, -\frac{1}{2}\openone_{})$, i.e.\
all eigenvalues of $J$ are non-vanishing. The potential~$Q$ will have block-matrix structure compatible with the
one of $J$.

\subsection*{Acknowledgements}

The authors have the pleasure to thank Prof. Adrian Constantin, Prof. Nikolay Kostov, Prof. Alexander Mikhailov and Dr. Rossen Ivanov for numerous useful discussions. Part of the work was done during authors visit at the Erwin Schr\"odinger International Institute for Mathematical Physics in the framework of the research programme ``Recent Advances in Integrable Systems of Hydrodynamic Type''  (October 2009). One of us (GGG) is grateful to the
organizers of the Kyiv conference for their hospitality. This material is based upon works supported by the Science Foundation of
Ireland (SFI), under Grant No. 09/RFP/MTH2144. We thank three anonymous referees for numerous useful suggestions.

\pdfbookmark[1]{References}{ref}
\LastPageEnding

\end{document}